\documentclass[]{aa}

\usepackage{graphicx}
\usepackage{txfonts}
\usepackage{xspace}
\usepackage{svg}
\usepackage{adjustbox}
\usepackage{listings}
\usepackage[frozencache,cachedir=minted-cache]{minted}

\newcommand{\snguess}{\texttt{SNGuess}\xspace}
\newcommand{\follow}{\texttt{FollowMe}\xspace}
\newcommand{\final}{\texttt{FinalBet}\xspace}
\newcommand{\elasticc}{\texttt{ELAsTiCC}\xspace}
\newcommand{\ampel}{\texttt{AMPEL}\xspace}
\newcommand{\desc}{\texttt{DESC}\xspace}
\newcommand{\parsnip}{\texttt{ParSNIP}\xspace}
\newcommand{\ulens}{$\mu$lens\xspace}

\newcommand{\newI}[1]{#1}   

\newcommand{\newR}[1]{#1}   

\usepackage[markup=underlined]{changes}

\begin{document}

   \title{AMPEL workflows for LSST: \newI{Modular and reproducible real-time photometric classification} }

\author{Jakob~Nordin\inst{\ref{inst1}}\thanks{jnordin@physik.hu-berlin.de}
 \and Valery~Brinnel\inst{\ref{inst1}}
 \and Jakob~van~Santen\inst{\ref{inst2}}
 \and Simeon Reusch\inst{\ref{inst1},\ref{inst2}}
 \and Marek Kowalski\inst{\ref{inst1},\ref{inst2}}
 }

\institute{
Institut fur Physik, Humboldt-Universität zu Berlin, D-12489 Berlin, Germany \label{inst1} 
\and 
Deutsches Elektronen Synchrotron DESY, Platanenallee 6, 15738 Zeuthen, Germany\label{inst2}
}

\titlerunning{AMPEL workflows for LSST}
\authorrunning{Nordin, Brinnel, van Santen et al}

\date{}


  \abstract
   {Modern time-domain astronomical surveys produce high throughput data streams which require tools for processing and analysis. This will be critical for programs making full use of the alert stream from the Vera Rubin Observatory (VRO), where spectroscopic labels will only be available for a small subset of all transients.}
   {We introduce how the AMPEL toolset can work as a code-to-data platform for the development of efficient, reproducible and flexible workflows for real-time astronomical application.
   }
   {
   The Extended LSST Astronomical Time-series Classification Challenge (ELAsTiCC) v1 dataset contains a wide range of simulated astronomical transients, taking both the expected VRO noise profile and cadence into account. We here introduce three different AMPEL channels constructed to highlight different uses of alert streams: to rapidly find infant transients (\snguess), to provide unbiased transient samples for follow-up (\follow) and to deliver final transient classifications (\final).
   These pipelines already contain placeholders for mechanisms which will be essential for the optimal usage of VRO alerts: combining different classifiers (built on boosted decision trees and deep neural networks), including host galaxy information, population priors and sampling non-gaussian photometric redshift distributions.
   }
   {
   All three channels are already working at a high level: \snguess correctly tags $\sim 99\%$ of all young supernovae, \follow illustrates how an unbiased subset of alerts can be selected for spectroscopic follow-up in the context of cosmological probes and \final includes priors to achieve successful classifications for $\gtrsim 80$\ \% of all extragalactic transients.
   }
   {
   Advanced statistical tools, including machine learning, will be critical for the next decade of real-time astronomy. However, training these models are only initial steps as the scientific application in a  real-time pipeline also relies on a long list of (conscious or unconscious) decisions: How should data be pre-filtered, probabilities combined, external information incorporated and thresholds set for reactions?
The fully functional workflows presented here are all public and can be used as starting points for any group wishing to optimize pipelines for their specific VRO science programs. \ampel is designed to allow this to be done in accordance with FAIR principles: both software and results can be easily shared and results reproduced. The code-to-data environment ensures that models developed this way can be directly applied to the real-time LSST stream parsed by \ampel.
   }


   \maketitle
%

\section{Introduction}

The torrent of real-time data delivered by astronomical observatories is constantly increasing. Current generation optical telescopes such as the Zwicky Transient Facility \citep[ZTF,][]{2018AJ....155....1G} already detect new astronomical transients at a rate so large that most are simply discarded; this rate will increase further with upcoming facilities such as the Vera C. Rubin Observatory \citep[VRO,][]{2019ApJ...873..111I}.

The futility of complete transient follow-up has long been apparent for science cases such as cosmology based on type Ia supernovae (SNe Ia) distances:
\cite{2013ApJ...763...88C} identified a sample of $752$ transients where the lightcurve alone suggested them to be SNe Ia. Continuing this trend, \cite{2017ApJ...843....6J} presented $\sim 1000$ photometric SNIa from the Pan-STARRS survey and \cite{2022MNRAS.514.5159M} $\sim 1400$ from the the Dark Energy Survey. 
The Legacy Survey of Space and Time (LSST), to be carried out at the VRO, is expected to provide samples up to $100 \times$ larger \citep{2009arXiv0912.0201L}.

Advanced statistical techniques, including Machine Learning (ML), are ideally suited for classification tasks, and several projects have already explored how to best carry out ML class inference based on photometric data. Examples of deep learning based models include
RAPID \citep{2019PASP..131k8002M},
PELICAN \citep{2019A&A...627A..21P},
SuperRAENN \citep{2020ApJ...905...94V},
SuperNNova \citep{2020MNRAS.491.4277M},
\parsnip \citep{2021AJ....162..275B} and SCONE \citep{2021AJ....162...67Q}.
Limits on the available labeled training samples force these to rely either on unsupervised learning or simulated datasets.
Classifiers based on extracted features, followed for example by decision trees, often require smaller training samples and have been similarly explored \citep{2016ApJS..225...31L,
2020ApJ...905...93H}. \cite{2021AJ....161..141S} and \cite{2022A&A...665A..99M} both train feature based classifiers directly on real alerts distributed by ZTF, closely resembling the data distribution of LSST.

These models typically classify light curves with a purity ranging from $50$ to $90$ \%, with most of the variability driven by the class being investigated. For example, distinguishing type Ibc supernovae (SN Ibc) from SNe Ia can be intrinsically difficult, while most superluminous supernovae (SLSNe) can be identified by their slow evolution and large intrinsic brightness. The different techniques have widely varying requirements in terms of input data, execution time and provided output. The models often yield systematically different results and might be more or less well suited for a particular application \citep{2023ApJS..267...25H}.
As the ML field is itself also rapidly changing, it is very unlikely that a single, stable  classification model will dominate the era of LSST. Rather, we can expect a constantly evolving zoo of models, out of which science projects can choose a combination best suited for their goals.

The critical question might thus not be which ML model to use, as this will change with time and science goal, but rather how to flexibly and systematically apply evolving algorithms to large data streams. Expanding on this issue leads to questions such as:
\begin{itemize}
    \item How can ML models be fairly compared, for a specific goal?
    \item How should a science user interested in one particular kind of (hypothetical?) transient design a real-time follow-up program?
    \item How can a science group apply an improved ML model to high throughput data streams?
    \item How can provenance -- knowing which model was used, by whom, for which classification, at what time, based on what data -- be guaranteed?
    \item How can archival or simulated data be systematically reprocessed and analyzed with new models?
    \item How can the \newI{statistical and systematic uncertainties} of a particular model be systematically evaluated and propagated to a final analysis?
    \item How can ML architectures developed by software domain experts be distributed to astronomers for active use?
    \item How can the real-time analysis and follow-up program carried out be compared with final, reprocessed data?
\end{itemize}

These questions all point to the need for frameworks capable of efficiently (re)processing datasets of various origins and systematically applying new statistical models while keeping track of both input configuration, runtime metadata, and output.

We here present \ampel, a modular framework designed for the analysis of time-domain data, which can be systematically used to answer the questions listed above while being efficient enough to parse large data volumes. \ampel will provide one of the (up to) seven access points for the VRO alert steam (the others being  Alerce, Babamul, Antares, Fink, Lasair and PittGoogle).
The outline of the paper is as follows:
Section~\ref{sec:ampel} provides an initial overview of the \ampel system. After this, the first Extended LSST Astronomical Time-series Classification Challenge (\elasticc) is introduced. Section~\ref{sec:pipe} describes three sample science cases designed to map expected uses of the VRO alert stream, and how ML based workflows can be implemented in \ampel. 
Section~\ref{sec:performance} evaluates their performance. 
Instructions for running a classification job locally are provided in section~\ref{sec:running}. We conclude in section~\ref{sec:conc}. The appendices contain a listing of the units that work as building blocks for the workflow (\ref{sec:units}), material from the \desc blind test (\ref{sec:blind}), an \ampel jobfile (\ref{sec:jobfile}) and sample examples of how data is stored in the \ampel database (\ref{sec:dbentries}).

\section{The AMPEL framework}
\label{sec:ampel}

AMPEL is a modular and scalable cross-platform framework with explicit provenance tracking,
suited for systematically processing large – possibly complex and heterogeneous – datasets \citep{2019A&A...631A.147N}. This includes analyzing, selecting, combining, updating, enriching and
reacting to data. Data sources can be both real-time data streams as well as archives of past events.
Although primarily developed to solve challenges in time-domain astrophysics, AMPEL is general enough to be utilized in various other fields.


\subsection{Structure}

AMPEL supports two modes of operation:
\begin{itemize}
\item A continuous mode is used for multi-user live systems, where the framework continuously
processes data streams and triggers external reactions where so configured. Multiple science programs are executed in parallel, and both data storage and algorthmic operations are efficiently deduplicated while keeping track of individual user data access rights. Data and results of a science user is identified by their unqiue \emph{channel} label. Following the VOEvent semantics \citep{2011ivoa.spec.0711S}, channels specify when and how, which unit(s) should be run on what data. 
\item A batch mode via the job system allows AMPEL to run similarly to the way traditional pipelines process data. The definition of a job \emph{schema} is then required, again specfiying the requested units. \ampel is lightweight enough to run on a laptop, providing a convenient local development environment. A job schema can be distributed (as a \texttt{yaml} file) and reexecuted by other users, either for development, reproduction or large scale execution at a data center.  
\end{itemize}

The two modes frequently integrate during a full analysis: an analysis schema is initially developed against a local \ampel installation based on smaller sets of training data, rerun at a computer center to reprocess a full set of archival data and finally converted into a channel for real-time data processing hosted by a live \ampel instance.

\ampel is a fully modular framework, thereby making it flexible, easily extendable and
maintainable. The core infrastructure executes pieces of code implemented in the form of \emph{units} according to a schema, thereby allowing contributed
code to be called directly during data processing. \ampel units are implemented as \texttt{python} modules, allowing popular astronomy libraries to be directly used.
Each unit belongs to one of four execution layers --- also called ``tiers'' --- that replace a
traditional pipeline architecture (see table~\ref{tab:tiers}). Each tier is independently scheduled, allowing
straightforward parallelization. Information exchange between tiers occurs via a dedicated
database.

\begin{table*}[t]
\centering
\begin{tabular}{ | l|l | }
            \hline
            Tier &  Operation \\
            \hline
            T0 & Filters large alert stream, selecting only transients of interest. \\
            T1 & Combines alert history with existing database or combine transient
information of different instruments. \\
            T2 & Evaluates transient properties (light-curve fitting, typing...). \\
            T3 & Estimates sample properties and reacts (ex: assign priorities to transients for
spectroscopic triggers). \\
 \hline
\end{tabular}
    \caption{\ampel operations are divided into four exection layers, each for a specific kind of task.}
    \label{tab:tiers}
\end{table*}

A standardization of AMPEL units behavior is enforced via abstraction (a contract on agreed
behavior between a parent class and its subclasses), enabled via AMPEL’s own abstract base
class, which provides some functionality lacking in the abstract base class provided by python.
In particular, it guaranties the arguments of implemented methods, which must match the
ones defined in abstract methods. By implementing public abstractions and thus adhering to
the AMPEL standards, units can be integrated in AMPEL workflows, which are automatically
managed by the infrastructure. The tier structrue provides an example of this: each implemented unit inherits data access methods from an execution layer base class. This allows \ampel to provide each unit with data in the format it expects: alerts (T0), series of datapoints (T1), a transient state (T2) or an ensemble of transients (T3).

The modular and heterogenous nature of \ampel workflows means that no common visualization or interactive front-end exists. Results are either stored locally for further exploration or propagated to dedicated front-end systems such as SkyPortal \citep{skyportal2019} or those hosted by other brokers. 
\newI{Instructions for running the workflows presented here can be found in section~\ref{sec:running}} and
further running and contact information can be found on the \ampel entry web page \footnote{\url{https://ampelproject.github.io/ampelastro/}} .

\subsection{Reproducibility, provenance and workflow management}
\label{sec:provenance}


%

The  scientific principle of absolute reproducibility cannot generally be achieved in time-domain astronomy, as we cannot influence when the Universe will again produce a rare or even unique transient event. This makes questions of provenance critical: is it possible to trace when an event was discovered, how it was interpreted and what actions were taken? This task gets continuously harder as detection rates, number of observational facilities and analysis complexities all increase. At the same time, this knowledge will be critical as we try to extrapolate from a small set of well-observed events unto the Universe as a whole.
Cosmology based on \newI{incomplete} samples of SNe Ia provides one example, but similar questions will appear
%
in any study wishing to make use of an assumed distribution of transients, for example going from the number of observed kilonovae to the amount of heavy nuclei predicted by an explosion model. This requires an understanding of what fraction of events, with characteristics as predicted by the model, would have been found by the observational program.
The era of photometric classification will add an additional and much more significant set of potential biases to such questions.

Provenance exists at multiple levels and with different purposes: from the capability to rerun some specific software element, to the precise logging of what happened (or not) at a specific time, to carrying out a consistent reanalysis of archived/simulated datasets.
\ampel was designed from the start with provenance and reproducibility in mind. The \ampel provenance structure is shown in figure~\ref{fig:prov} and described below, while a sample jobfile can be found in appendix~\ref{sec:jobfile} and sample database entries containing provenance fields in appendix~\ref{sec:dbentries}.

\begin{figure*}
\center{
\includesvg[inkscapelatex=false, width=0.7 \textwidth]{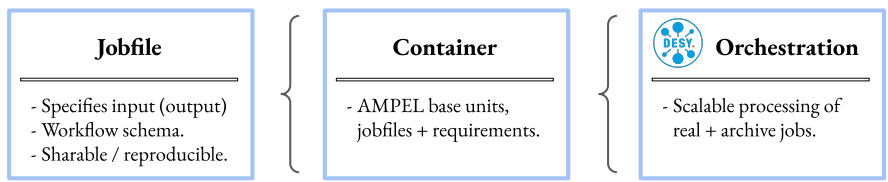}
}
\caption{Schematics of provenance and management features. \emph{Top section:} Relationship between measurements and derived products stored in the \ampel DB. \emph{Mid:} Additional DB collections designed to guarantee provenance. \emph{Bottom:} Workflow reproduction at different levels: as jobfile, container and orchestrated live instance.}
\label{fig:prov}
\end{figure*}

The top section of figure~\ref{fig:prov} illustrates the accumulation of data within \ampel: \emph{Measurements} are the raw data elements obtained from telescopes, simulations or external archives. Measurements stored in \ampel are considered immutable and are never replaced or edited. What information is available to a user at a specific time is instead recored by a \emph{state}, at its core a collection of datapoint references. An updated measurement thus translates into the addition of a new measurement being added and a new state created which has the references to the superceded datapoint replaced with that of the newly created. In this way it is straightforward to keep track of how the available information grows with time as well as manage users with different data access. Finally, \emph{augmentations} to the data, or derivations made by units in the \ampel system (such as photometric classifications) are always tied directly to the state or datapoint used to derive them. This first level of provenance thus makes it possible to track the evolution of knowledge.

The mid section of Fig~\ref{fig:prov} illustrates the additional database collections automatically generated by \ampel to ensure that it is known what was done at which time, and which software methods were used to produce results. The \emph{stock} collection contains records of all events considered as unique objects, by any observer. Each other datapoint and/or measurement contains references to the objects they are linked to through the stock id. Furthermore, the stock \emph{journal} field lists all \ampel processing activities associated with an object: when the object was added as well as when subsequent (T2) operations were completed. Each journal entry includes references to the software modules which completed the operation through the \emph{traceid}. The \emph{log} collection records any log messages produced by any of the executed \ampel units through the specific \texttt{AmpelLogger} class which is provided to each logical unit. The \emph{conf} collection stores the full configurations of all units executed by the \ampel instance. Finally, the \emph{trace} collection explicitly connects which software version, system environment and configuration which was used to carry out an operation.

The final section of figure~\ref{fig:prov} highlights the workflow structure. The core component here is the \emph{jobfile} which lists the series of units which are to be executed. These will specify both the source of data, how this should be analyzed as well as what reactions should be taken. A sample jobfile for processing \elasticc alerts along the pipelines discussed below is further discribed in appendix~\ref{sec:jobfile}. A jobfile can be executed by anyone with an \ampel environment, and (for example) included in a publication for easy result reproduction. The jobfile, together with the specific \ampel environment can be stored in a container, either for storage or for mounting at a computer center for live processing. An example of the latter is the live, public instance hosted by the DESY Zeuthen computer center, which processes both live ZTF alerts as well as the \desc simulated streams (and eventually one of the LSST alert streams).

Together, these components create a system which enables users to easily assemble real-time workflows which are guaranteed to be reproducible without having to specifically design their algorithms for this: everything from results to how and when they were determined is recorded by the system.

\section{The ELAsTiCC challenge}
\label{sec:elasticc}

The Dark Energy Science Collaboration (DESC) initiated ELAsTiCC ("Extended LSST Astronomical Time-series Classification Challenge") in order to to test broker infrastructures and to challenge photometric classification algorithms, with the additional goal of preparing the community for how to best use the LSST data to measure cosmological parameters.\footnote{\url{https://portal.nersc.gov/cfs/lsst/DESC_TD_PUBLIC/ELASTICC/}} Two ELAsTiCC blind tests have so far been carried out: v1 in 2022 and v2 in 2023/2024. We here discuss methods and results developed for v1, but a similar setup was also used for v2. 

The ELAsTiCC simulation used \texttt{SNANA} \citep{2009PASP..121.1028K} to generate lightcurves for $15$ kinds of non-recurrent transients, five kinds of recurrent periodic transients (variable stars) and one kind of non-periodic reccurrent transients (AGNs). The non-recurrent transients were further divided into \emph{SN-like} (SNIa, SNIb/c, SNII, SNIax, SN91bg), \emph{Fast} (Kilonovae, M-dwarf Flares, Dwarf Novae, micro-lens events - \ulens) and \emph{Long} (superluminous supernovae - SNLS, tidal disruption events - TDE, intermediate luminosity optical transients - ILOT, calcium rich transients - CART, pair instabilitiy supernovae - PISN).\footnote{\url{https://github.com/LSSTDESC/elasticc/blob/ee51c2e8dd04128839813d4bee2c12f19410c060/taxonomy/taxonomy.ipynb}}
The \elasticc simulation also provides spectroscopic and/or photometric redshift estimates as well as host galaxy information, modeled after what is expected to be available in LSST alerts.
Knowing the redshift and thus distance to the transient greatly increases the probability of a correct classification and follow-up decisions \newI{as the absolute brightness can be included among the classification features, \emph{if so wished}}. While spectroscopic redshifts will only exist for a small subset of all galaxies, photometric redshift estimates will be generally available, albeit with large uncertainties. Furthermore, several of the extragalactic transient classes are predominantly found in particular host galaxy environments. In practice, using such information comes at a risk: the host galaxy association might be spurious, or the photometric redshift catastrophically wrong. The assumption of transient-environment correlations carries with it a risk of biased selection functions.
%
\newR{The ELAsTiCC simulation derives several transient class models from the previous PLAsTiCC classification challenge, references and data for these are conveniently available.\footnote{\url{https://doi.org/10.5281/zenodo.6672739}}
}

Training samples with labels were distributed for each simulated class, divided into $40$ batches each.\footnote{\url{https://portal.nersc.gov/cfs/lsst/DESC\_TD\_PUBLIC/ELASTICC/TRAINING_SAMPLES/}}
A test sample consisting of $5$ million transients were simulated as a blinded test. The lightcurves were transformed into 50 million alerts wich were distributed using a Kafka server during a 6 month period, without revealing the true transient class. The setup corresponded to a simulated slice of the sky as this would be observed by LSST during the 10 year survey time. Each of the VRO brokers was invited to submit classification reports which were compared with the truth tables maintained by DESC.

\section{The \ampel workflow design for ELAsTiCC}
\label{sec:pipe}

\subsection{Sample science cases}

VRO was designed for a wide range of different science applications, many of which will not make use of its real-time products at all. However, for many time-domain programs the short reaction times enabled by real-time alerts will be critical in terms of catching fast variability and understanding physical processes. This includes searches for multi-messenger counterparts, reactions to newborn or rapidly varying transients, searches for exotic phenomena or more generally, any program which might wish to obtain follow-up observations of transient phenomena.
In practice, it might be advantageous for any program wishing to make use of the time-domain aspect of LSST data to make use of the alert distribution for convenience, reproducibility and collaboration with other science groups.

The three \elasticc workflows presented here (\snguess, \follow and \final) were designed to highlight different ways in which ML models can be put to use.
They focus on non-recurrent extragalactic transients; no effort was made to identify galactic recurrent transients.
\newI{\follow and \final both add a deep learning classification stage to the \snguess workflow, but apply different criteria and priors.}
The primary goal at this stage was not to produce optimal classification metrics, but rather present example workflows that have enough flexibility that different science groups can develop and optimize these for specific science goals. 

\paragraph{\snguess} Some of the most exciting real-time programs aim at finding newly born transients.
The first hours to minutes carry information that will later be lost, and both rapid identification and follow-up are needed. The deep LSST survey will allow the detection of extremely young supernovae in nearby galaxies as well as fainter transients like kilonovae at larger distances than previously possible. As so little information is available at the first detection, it is impossible to achieve a full classification, and the most critical real-time question becomes "Is this a young extragalactic transient?" The \snguess program was designed to answer this question. The desired outcome here is binary --- good/bad.  The primary training goal is to reject as many variable stars and AGNs as possible, while retaining any potential infant transients. A full program might couple this output with other constraints, for example from multi-messenger confidence areas or the availability of follow-up facilities.

\paragraph{\follow} A large fraction of all time-domain VRO programs will have to rely on photometric classification schemes. These classifications will be derived based on complete lightcurves released after the transient has faded, but any complete scientific program needs to include follow-up observation of a subset of all identified candidates in order to obtain labels to train and verify the performance of the classifier. The labeled subset (obtained while transients were active) should ideally provide unbiased testing coverage of all relevant properties of the final photometric samples.
For this purpose, \cite{2019MNRAS.483....2I} presents an active learning methodology aimed at continously finding the best active transients to complement the training sample. Alternatively, methods such as STACCATO try to reweight the available training samples \newI{such that events in less populated areas of the parameter space (e.g. fainter) are assigned a higher weight}
\citep{2018MNRAS.473.3969R}. \newI{The goal of both approaches is to create a final model with minimal classification errors.}
\newR{\follow takes a slightly different path and accepts that photometric classifications have an unavoidable uncertainty. The goal here is rather to obtain the follow-up labels needed to accurately model this statistical error. As obtaining labels is expensive, we wish this process to be as efficient as possible and that labels are primarily obtained for cases with uncertain classification.} The problem of how to separate SNe Ia (which are excellent cosmological standard candles) from SNe Ibc (which bias distance estimates if mistaken for SNe Ia) is taken as a sample science goal, and the objective is to collect an unbiased and observable selection of transients that are likely to be misclassified, spanning both SN and environment variability. 

\paragraph{\final}
The eventual goal of any transient program will be to obtain ``final'' classifications that incorporate all available information, including probabilistic estimates derived in parallel or prior to the development of the alert classifier.
Even if such a final classification is derived long after a trasient has faded, it will be critical for sample provenance that there is a fully traceable and continuous link between the decisions made in real time and the post-processing carried out later. Only with this information will it be possible to distinguish the effects of photometric calibration, classification pipeline and any applied priors.
The \final classification channel showcases the impact of the inclusion of three priors based on the transient redshift, expected transient rates and host galaxy ($u-g$) color. The shape of these are, to various degrees, arbitrary but can be easily adjusted.

\begin{figure}
\center{
\includesvg[inkscapelatex=false, width=0.9 \columnwidth]{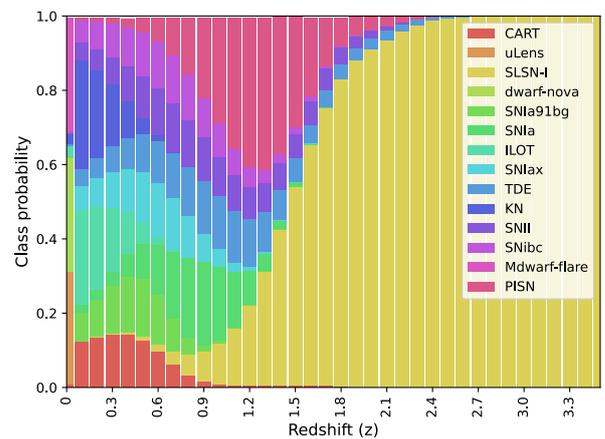}
}
\caption{Redshift priors as applied by the \final classifier channel. The best fit transient redshift (x-axis) will be used to derive the probability prior to add to the classifier output. For example: Transients at $z=2$ are assumed to be $\approx 90$ \% SLSN, with TDE, SNII, PISN also being possible. }
\label{fig:zprior}
\end{figure}

The redshift prior is derived from the redshift distribution of each class as found in the \elasticc training sample (batch $1$ to $20$). This is integrated into redshift bins of width $0.1$, and the distribution for each class is normalized to unit area. The contribution to each redshift bin is then summed for all transient classes and renormalized. The resulting redshift weight is shown in Fig.~ \ref{fig:zprior}. No consideration is taken of the relative rates of transients nor of the shape of the distribution. The main effective shift is that (assumed) high redshift events will be heavily weighted towards PISN or SLSN. Note that the training sample is expected to differ from the test sample, and that this prior is clearly suboptimal.

\begin{figure}
\center{
\includesvg[inkscapelatex=false, width=0.9 \columnwidth]{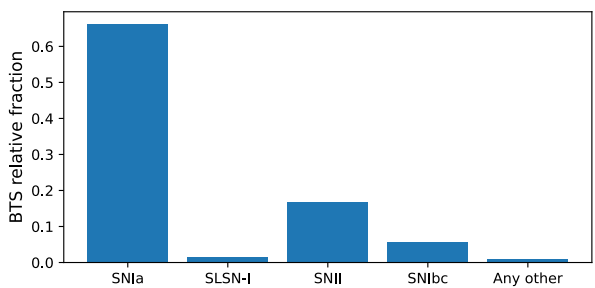}
}
\caption{Relative probabilities of classes as estimated from BTS. Each class (also these not defined by BTS) has a minimal probability of $0.01$ assigned.}
\label{fig:btsprior}
\end{figure}

The rate prior is directly taken from the ZTF Bright Transient Survey \citep[(BTS)]{2020ApJ...895...32F}. Figure\  \ref{fig:btsprior} shows the relative rates of SNIa, SLSN, SNII and SNIbc.
All other classes are assumed to have a relative rate of $1$ \% (some were found at even lower rates in BTS, some did not exist at all). The main effective shift of this prior is to make SNIa and SNII much more likely predictions. 
\newR{
The rates actually used in \elasticc were not known at the time of the challenge, but have since been made public through the simulation files.\footnote{\url{https://github.com/LSSTDESC/elasticc/tree/main/model_config}} The accuracy of the priors thus correspond to a realistic scenario where some knowledge of the expected rates exist, but which does perfectly match real values.
}

Finally, the $u-g$ color derived from the host galaxy photometry provided in the alerts (where available) is used to add a host prior. In this case, a normal distribution was fit to the $u-g$ color distributions of each class in the training sample (see Table~\ref{tab:ug}). When evaluating an alert, the gaussian probability of the alert host ($u-g$) color is determined for each class. These are used to reweight the class probabilities, after normalizing the sum to one.

\begin{table}[]
\centering
\begin{tabular}{ l|c|c }
            \hline
            Class &  Mean (u-g) &  Std (u-g) \\
            \hline
            CART     & 1.17 & 0.55 \\
            ILOT     & 1.15 & 0.51 \\
            KN       & 1.31 & 0.33 \\
            PISN     & 0.34 & 0.45 \\
            SLSN-I   & 0.38 & 0.37 \\
            SNIa91bg & 1.89 & 0.43 \\
            SNIa     & 0.61 & 0.39 \\
            SNIax    & 0.69 & 0.37 \\
            SNIbc    & 0.95 & 0.57 \\
            SNII     & 0.88 & 0.53 \\
            TDE      & 0.88 & 0.72 \\
 \hline
\end{tabular}
    \caption{Host galaxy color fits for training sample. A normal distribution fit to the host ($u-g$) colors of the training sample yields the location and scale provided here.}
    \label{tab:ug}
\end{table}

The \final class will reweight the probabilities derived from the alert data in \follow based on the three  priors discussed above. While it was unclear during the design phase whether these particular choice of priors would improve classification results, carefully designed priors will improve the classification performance at the cost of a potentially biased selection.

\subsection{Selection and training of classification models}
\label{sec:training}

Two ML models are used by workflows presented here. \snguess is based on the gradient boosting decision tree algorithm implemented in \texttt{XGBoost} \citep{2016arXiv160302754C}, where the features used and the training steps match those described in \cite{2022A&A...665A..99M}. Batches $1$ to $20$ were used as labeled input data. The training steps include k-fold sampled grid search to determine model parameters and can be replicated using the \texttt{elasticc2} package.\footnote{https://github.com/desy-multimessenger/elasticc2} The \ampel units \texttt{T2RiseDeclineStat} and \texttt{T2MultiXGBClassifier} are used to carry out classifications based on the trained models.

\follow and \final attempts at making a full classification of non-recurrent transients, as selected by \snguess, using the deep learning \parsnip model. If the initial \snguess stage evaluates a transient as a recurrent event no further classification is attempted. 
\parsnip is a hybrid generative model which combines explicit and intrinsic latent variables \citep{2021AJ....162..275B}. The first kind is modeled using known physical relations (redshift, for example) while the second are determined using a modified version of a variational autoencoder. \parsnip effectively allows for one encoder-decoder to work across distances, bandpasses and noise levels.
The explicit variables of the \parsnip model include a well defined reference (peak) time for each transient, making it less suited for recurring transients. Finally, the gradient boosted decision tree package \texttt{lightgbm} is used to train a classifier based on the estimated transient variables.
The model used in the \follow and \final classifiers was derived using the \texttt{parsnip\_train} and \texttt{parsnip\_predict} methods, based on the $1$ to $20$ batch samples of non-recurrent transients and with a factor $10$ augmentation. 
The \parsnip inference is carried out by the \texttt{T2RunParsnip} unit.

We choose not to use \parsnip's option to include photometric redshifts, as we found it more efficient to rerun a fixed redshift model at a set of discrete redshifts.

\subsection{Final workflows}

\begin{figure}
   \centering
   \includegraphics[width=\hsize]{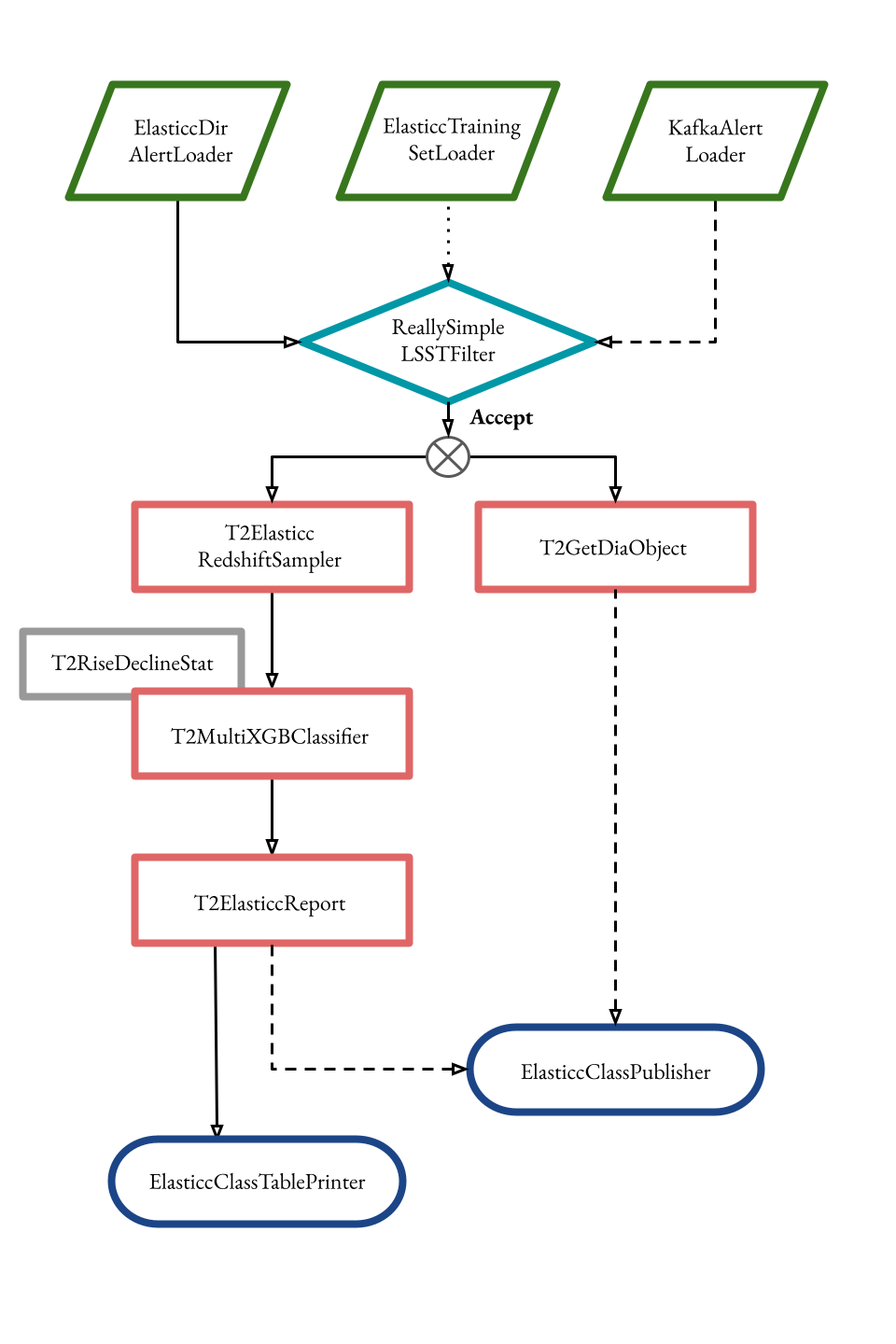}
      \caption{AMPEL schema for the \snguess channel. \emph{T0:} Top row (green parallelograms) show possible loader classes, each of which provides LSST like alerts to the alert filter (cyan diamond). \emph{T2:} Alerts which pass the filter stage are stored into the database and augment/compute operations (rectangles) scheduled for execution. \texttt{T2RiseDeclineState} is the base class for \texttt{T2MultiXGBClassifier}, responsible for the feature extraction. \emph{T3:} The output stage (blue ovals) reacts based on the operations, either through collecting data for local printout or for the submission of classification reports.
      Arrows show information flow. The solid line path shows the units included in the demonstration job file (section~\ref{sec:running} and appendix~\ref{sec:jobfile}). The dashed path corresponds to the schema executed during the \elasticc challenge.
              }
         \label{fig:snguessflow}
   \end{figure}

\begin{figure}
   \centering
  \includegraphics[width=\hsize]{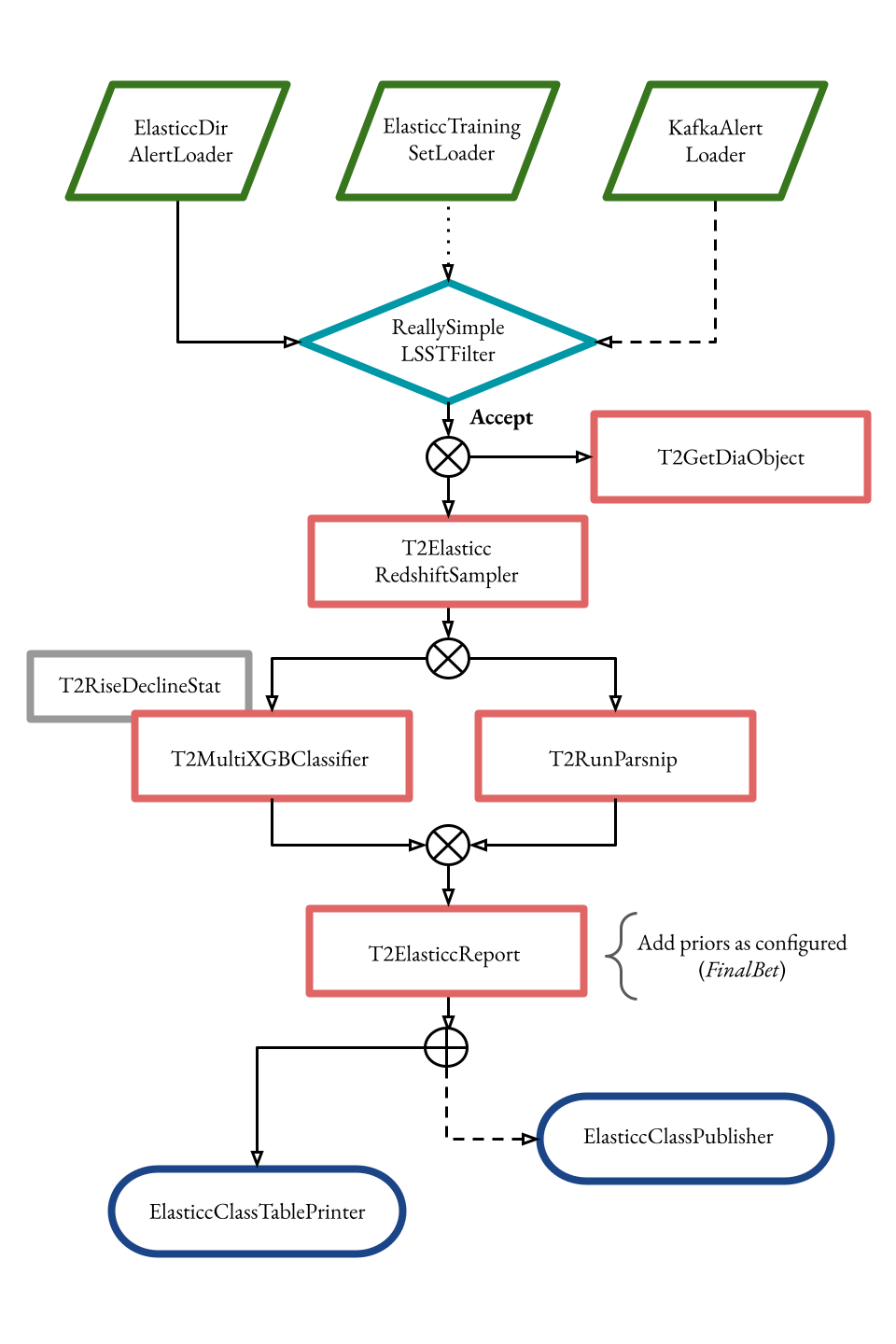}
      \caption{AMPEL schema for the \follow and \final channels. The \snguess workflow (Fig. \ref{fig:snguessflow}) is extended by the execution of the \texttt{T2RunParsnip} classification unit. The output of this is also forwarded to \texttt{T2ElasticcReport} and included in the combined classification. The \final channel will at this stage also incorporate priors. (The connection between \texttt{T2GetDiaObject} and \texttt{ElasticcClassPublisher} is suppressed for clarity.)
              }
         \label{fig:parsnipflow}
   \end{figure}

The workflows used by the \snguess, \follow and \final channels are illustrated in Figures \ref{fig:snguessflow} and \ref{fig:parsnipflow}. Each box corresponds to an individual \ampel unit implemented as a Python class; its shape and color denoting the tier it belongs to. The purpose and logic of each unit is summarized in appendix~\ref{sec:units}, with the full implementation available in the Ampel-ZTF\footnote{https://github.com/AmpelProject/Ampel-ZTF} or Ampel-HU-astro\footnote{https://github.com/AmpelProject/Ampel-HU-astro} repositories (in a \texttt{Python} file with the same name as the unit). The core tasks consist in running the classsifier introduced above, but two additional steps were needed to account for photometric redshifts and potential negative detections. These are summarized here, with more information in the appendix.  

ELAsTiCC alerts can contain different levels of host galaxy information: spectroscopic, photometric or no redshift at all. Furthermore, there can be more than one potential host galaxy. This will be true of real LSST alerts as well. The \parsnip classifier implemented in  \texttt{T2RunParsnip} expects a series of redshifts, and will then carry out fits at each of these discrete values. The final classification consists of the average of these values, weighted both by an input weight as well as the \newI{probability obtained when comparing the fit $\chi^2$ value with a $\chi$ distribution}.
\newI{"Catastrophic" photometric redshifts, caused by a bimodal likelihood with some probability of getting completely wrong values}, can thus potentially be overcome by a much better fit likelihood at the true redshift. Similarly, transients without any host information can be fit with a "flat" redshift prior. The \texttt{T2ElasticcRedshiftSampler} unit is responsible for providing this input set of redshifts, together with a set of weights. First, the most likely host is identified based on the distance and size. Second, three redshifts are sampled from the potentially non-gaussian uncertainty distribution with weights calculated based on the integrated probability. In case no redshift information exists \{$0.01, 0.2, 0.5, 0.8$\} will be tried. If a spectroscopic redshift exists this will be used.

The \snguess classifier was originally developed based on real ZTF alerts, aimed at identifying non-recurring supernovae events. However, recurring classes such as AGN or variable stars will frequently provide negative detections when reference images contain elevated flux levels \newR{and, while \elasticc realistically provides also negative flux values, the first version of \snguess used here could not handle these}. The following extra steps were thus taken in order to process negative flux alerts: Alerts with negative detections and  $z>0.001$ will be labeled as AGN and alerts with negative detections and $z<0.001$ will be lableed as \ulens events (more than $20\%$ of all photometric datapoints being negative) or eclipsing binary events (less than $20\%$ negative). 
\newR{These fixed, semi-arbitrary limits were set during initial training based on visual inspection of a subset of training alerts. Later versions of \snguess, used post-\elasticc, were modified to handle negative flux values and do not use these arbitrary limits (consequently further improving performance). }

\section{Classifier performance}
\label{sec:performance}

As discussed above, only half of the $40$ training sample batches were used for training.
%
We here test the three classification workflows based on training data that were set aside for this purpose: training batch $37$ for non-recurrent transient types, and batches $34$ to $37$ for recurrent events. The enlarged test sample for the latter types is necessary because while they have longer lightcurves, each batch contains fewer unique events. The test data used here is thus distinct from the training sample, but based on the same relative distributions.

The classification schema can label alerts as "Noise". In the context of \elasticc we know that all alerts correspond to a (simulated) astronomical transient, while the real data stream will likely be dominated by noise close to the detection threshold. Out of all alerts in the test sample, $2.3$ \% were labeled as "Noise". These were ignored for the purpose of the evaluations discussed here.



\subsection{\snguess}

\begin{figure}
   \centering
  \includegraphics[width=\hsize]{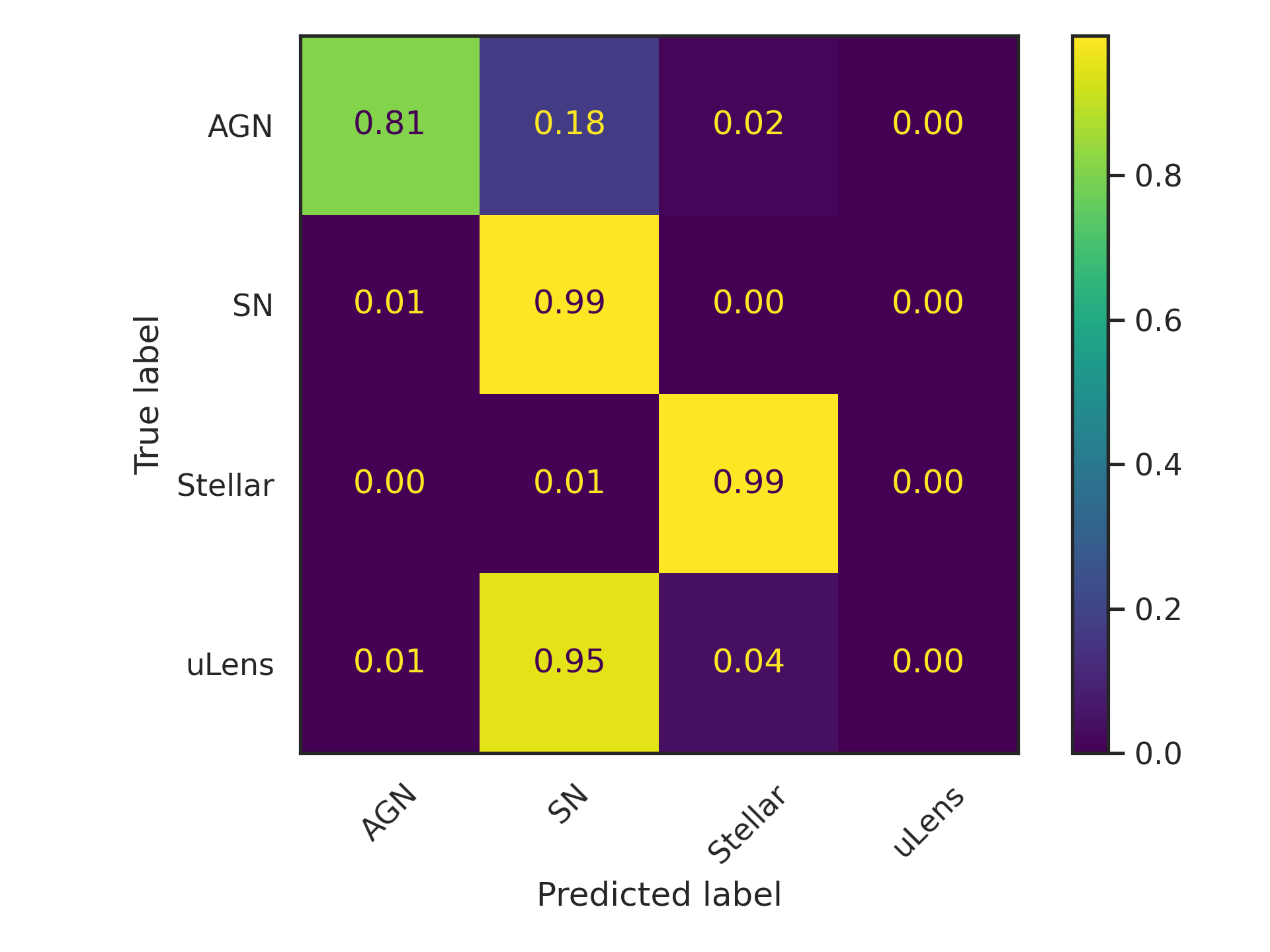}
      \caption{
Confusion matrix for alerts derived from \elasticc labeled sample \#34-\#37 (not included in training). Alerts were classified as either "AGN" (recurrent, extragalactic transients), "Stellar" (galactic transients) or "SN-like". The SN-like subset encompass non-recurrent extragalactic transients from the SN, Long and Fast taxonomy groups. The \ulens subset is shown separately, as the \snguess classifier was developed without including this subset. The \snguess classifier does an excellent job of distinguishing SN and Stellar events. The conservative approach can be seen in that a substantial fraction of all AGN and almost all \ulens events are tagged as SN-like \newI{when first detected}. }
         \label{fig:test_snguess}
\end{figure}

The goal of the \snguess workflow is to identify non-recurrent transients soon after first detection in order to make rapid follow-up observations possible, with an emphasis on not missing potential targets. To focus on the interesting subset, we only include alerts with $<=5$ detections; anything with more observations is assumed to be no longer interesting.
The \snguess confusion matrix for separating non-recurrent extragalactic transients, AGN, Stellar variables and \ulens alerts is shown in Fig.~\ref{fig:test_snguess}. The main goal of the analysis channel is achieved, as almost every target event is retained ($99$ \%), and a similarly large fraction of stellar variable events are rejected. However, a significant fraction ($18$ \%) of all AGNs would when first detected be tagged as non-recurrent. Such detections can easily be removed by additionally rejecting detections close to galaxy cores, but at a loss of other nuclear transients. 
Finally, as discussed above, the features used for \snguess were designed for positive detections and without considering \ulens events, which causes these events to also be classified as SNe.\footnote{An updated classifier version corrects for this, but was not used for \elasticc. In practice, \ulens events will be rare, and intrinsically interesting if discovered.}

The core \snguess workflow does not contain any constraints on target brightness, connection to potential multi-messenger signals or position relative to known galaxies. A full implementation would include such limits as well, which would further limit the sample and reduce AGN contamination.

\subsection{\follow}

The \follow workflow investigates the case where a selection should be made of which transients to follow up while they are still active. As a demonstration, we investigate the known problem of distinguishing SNe Ia from SN Ibc based on their lightcurve properties. This is likely to be one of the dominant astrophysical systematics for the LSST SN cosmology program.

This \newI{workflow evaluation}
thus only makes use of alerts generated by the SNIa and SNIbc models. These are furthermore restricted to those with a restframe duration (based on the photometric redshift) between $10$ and $30$ days and a last magnitude brighter than $22$ mag. 
These choices roughly mimic the constraints of a spectroscopic follow-program with some access to intermediate/large scale facilities.

\begin{figure}
   \centering
  \includegraphics[width=\hsize]{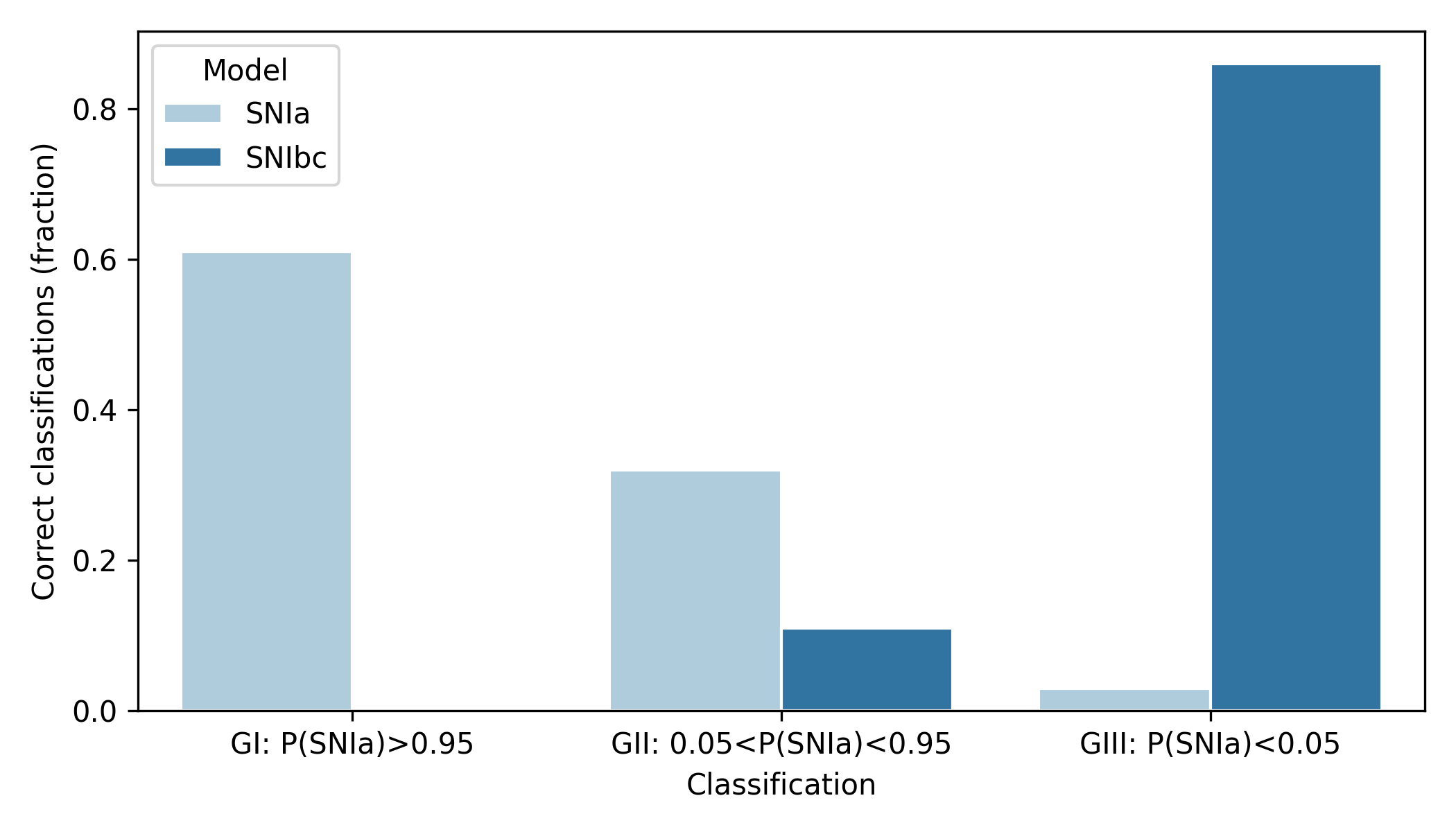}
  \includegraphics[width=\hsize]{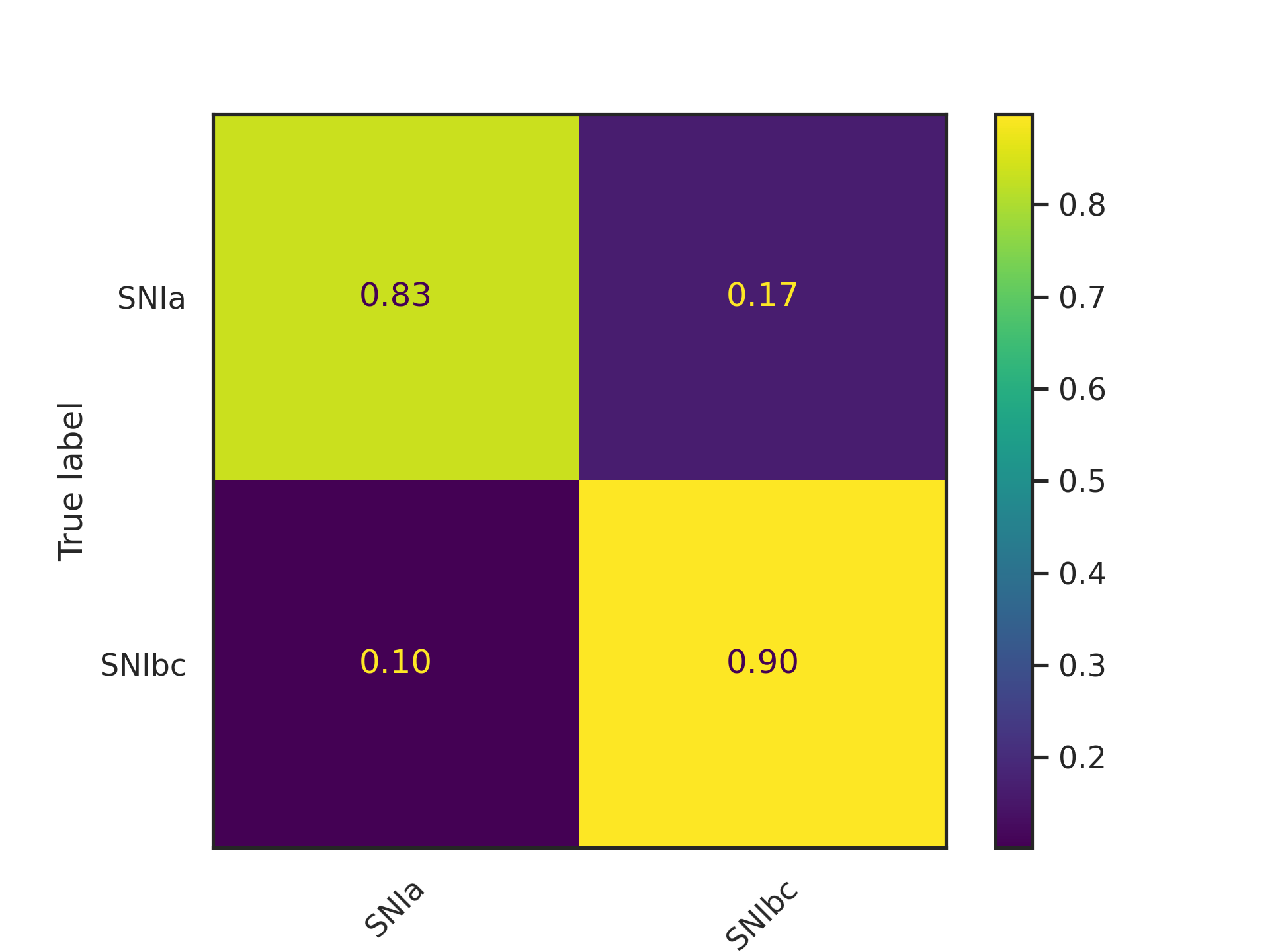}
      \caption{
\emph{Top:} Separation of SNIa and SNIbc transients into three groups (I, II, III) based on the SNIa probability reported by the \follow classifier. G-I with P(SNIa)>0.95 encompasses most SNIa events, while G-III captures almost all SNIbc but with a small SNIa contamination. G-II highlights the interesting area where the classifier does not confidently identify a transient as SNIa or SNIbc.
\emph{Bottom:} Confusion matrix for SNIa and SNIbc transients in G-II, based on the most probably transient kind reported. Also within this group, most transients are correctly classified. The residual uncertainty expressed here can be taken into account by statistical models for cosmological analysis, but an observational campaign would be needed to verify whether the reported uncertainty is correct.
              }
         \label{fig:test_follow}
\end{figure}

A fraction of all alerts will be classified with a high confidence by the photometric model. If this prediction is true, this (hopefully) large subset does \emph{not} need to be extensively verified by spectroscopic follow-up.
We investigate this assumption by asserting that alerts
with a probability of being a SNIa higher than $95$ \% or lower than $5$ \% are taken to have an accurate classification. As can be seen in Fig.~\ref{fig:test_follow}, $60$ \% of all SNe are correctly identified using these limits with a negligible contamination. More than $80$ \% of SNIbc alerts would be correctly rejected (P(SNIa)$<0.05$), but $3$ \% of SNe Ia alerts would be missed. Changing the rejection threshold such that no SNe Ia are lost would cause a large fraction of the SNIbc alerts to be retained as well.

We now focus on the central region of alerts with ambiguous classification, i.e. where no class has a probability larger than $95$\xspace\%. The confusion matrix \emph{only for this subset} is shown in the lower panel of Fig.~\ref{fig:test_follow}. While most lightcurves are correctly classified, there is a significant amount of misclassification ($10$ to $20$ \%). The key goal of a \follow workflow would be to sample a subset of transients from this uncertain set for spectroscopic identification in order to construct an accurate statistical model.  Such a model could, for example, be used to include photometrically-identified SNe Ia in cosmological studies. As this is a restricted subset designed to be observable, a systematic follow-up campaign might be feasible also for surveys such as LSST.

\begin{figure*}
   \centering
  \includegraphics[width=0.72 \hsize]{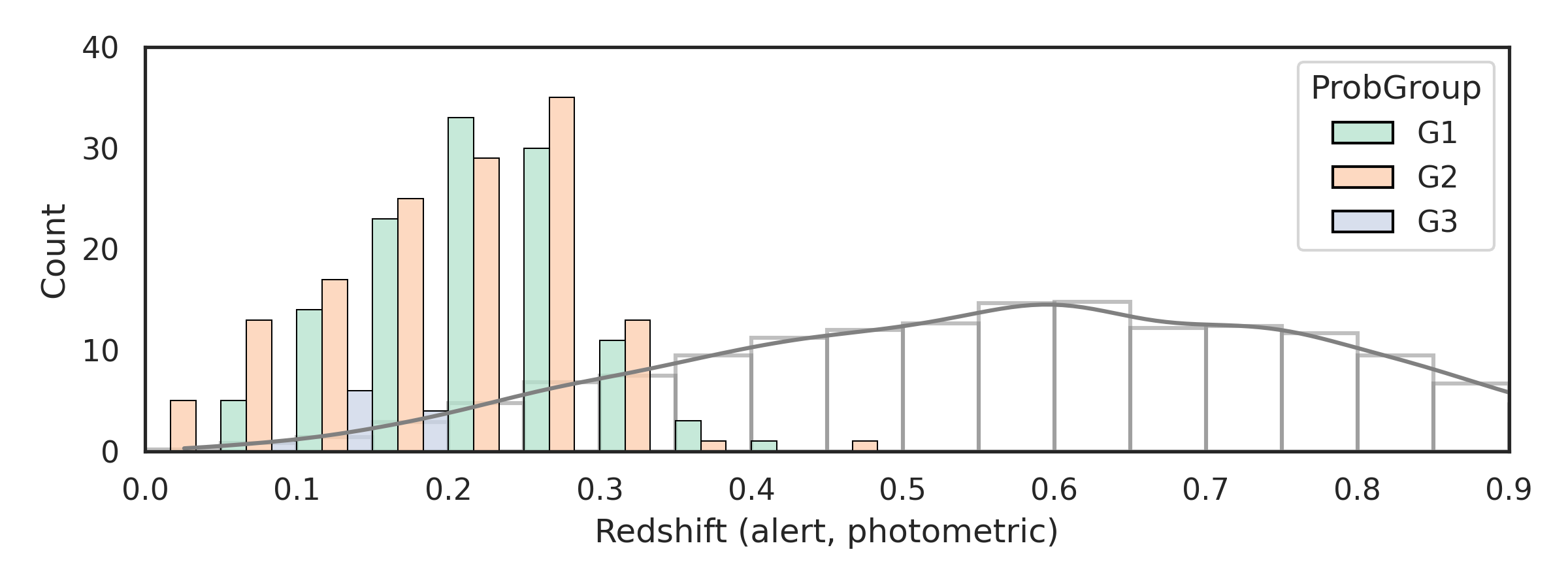}
  \includegraphics[width=0.27 \hsize]{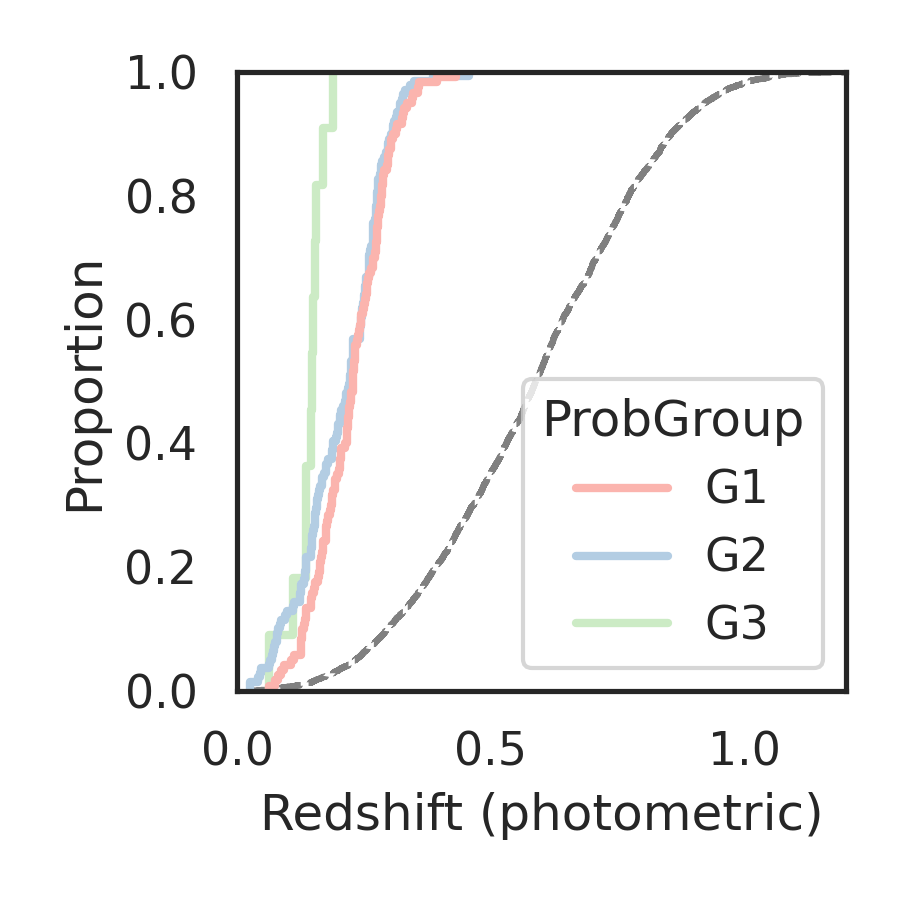}
  \includegraphics[width=0.72 \hsize]{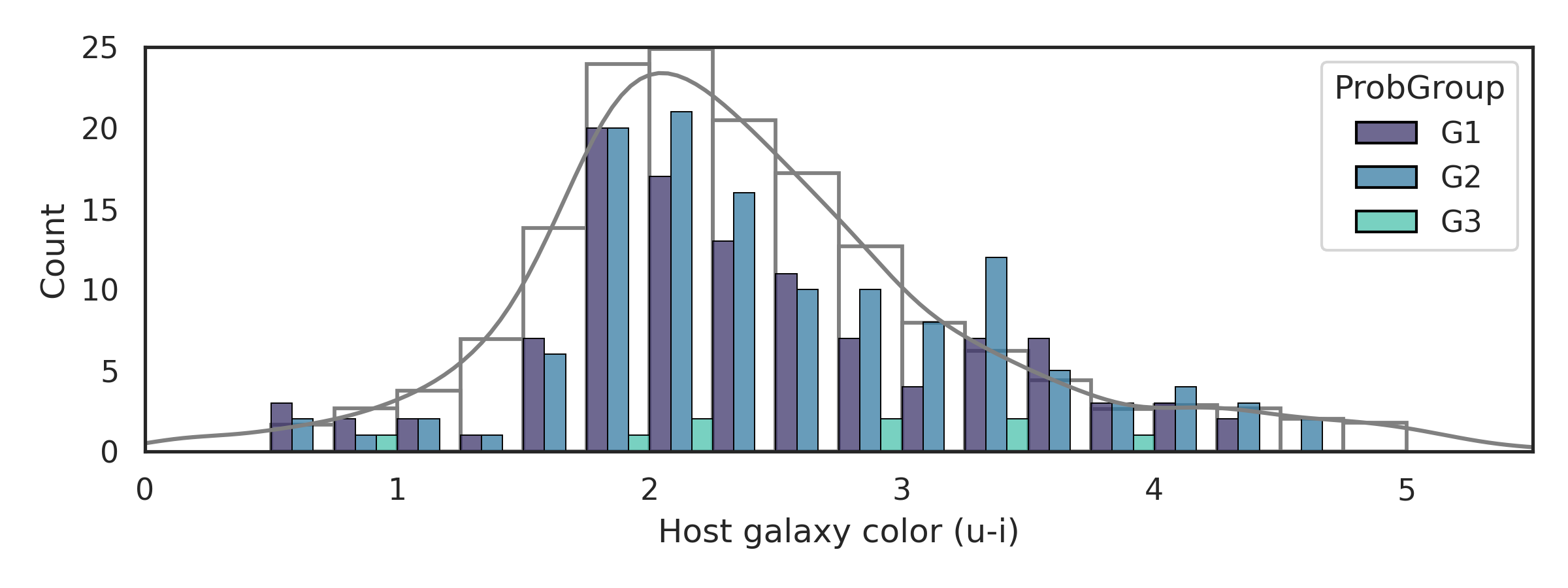}
  \includegraphics[width=0.27 \hsize]{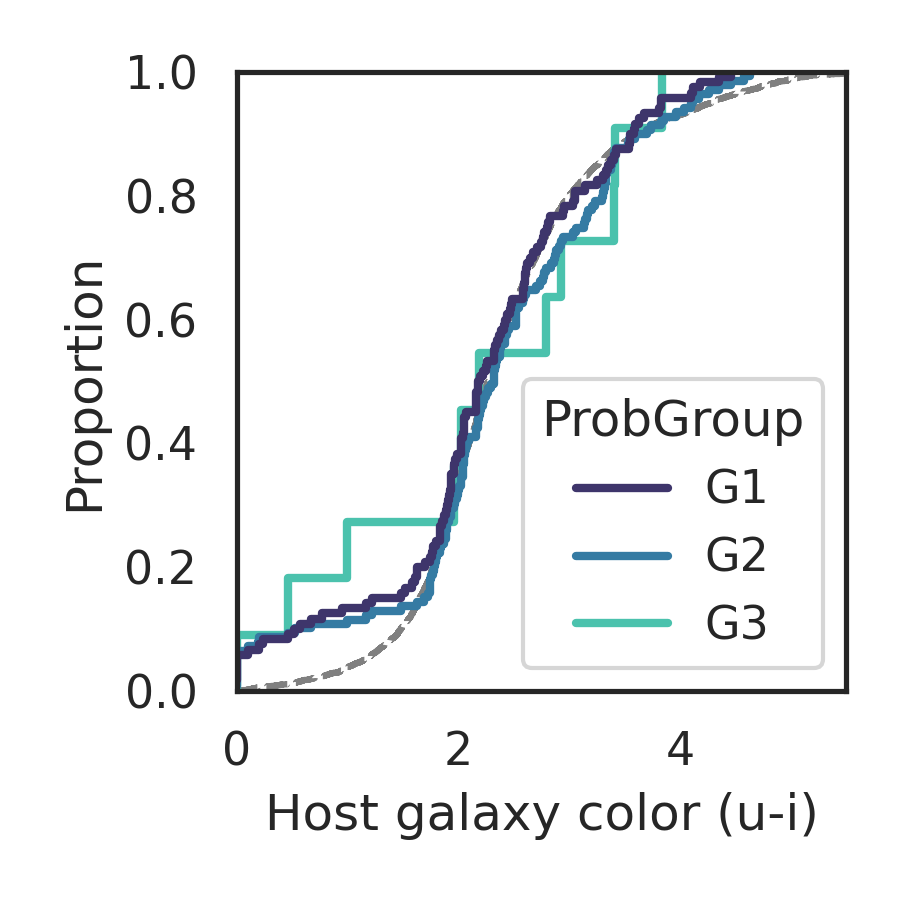}
  \caption{
        Distribution of photometric redshift (top) and u-i galaxy color (bottom) for SNe Ia in the \elasticc training sample. Each panel shows the distributions for "secure" SNIa (G-I), "secure" SNIbc (G-III) and uncertain (G-II), together with the full SNIa population (grey line). Right panels shows the cumuluative distributions.
              }
         \label{fig:test_follow_dist}
\end{figure*}

The critical question, however, will be whether this subset is be representative of the (unknown) parent SN sample it was drawn from.
A bias among objects chosen for labeling could lead to systematic errors in how the photometric sample is understood. The properties available for the \elasticc training sample do not allow for fully realistic studies, but we can demonstrate the process through the (photometric) redshift and u-i host galaxy color.
The distributions for these properties, together with the underlying distribution of the full SNIa training sample, are shown in Fig.\xspace\ref{fig:test_follow_dist}.
As expected, the redshift distribution for the selected SNIa sample is significantly different from that of the full training set --- a natural consequence of the magnitude limit imposed by the follow-up program. It can also be seen that the "secure SNIbc" subset (G-III) is concentrated at low redshift.
The galaxy color distributions, on the other hand, are similar between the different subsets and the underlying parent distribution.  The KS test p-value when comparing whether the true SNIa distribution differs from the mixed G-II alert group is a non-significant $0.22$.
This is an encouraging sign that the mixed subset here chosen for classification can produce a labeled test set that can be used to verify the classification across different \newI{host galaxy environments and supernova subtypes}.

\subsection{\final}

\begin{figure}
   \centering
  \includegraphics[width=\hsize]{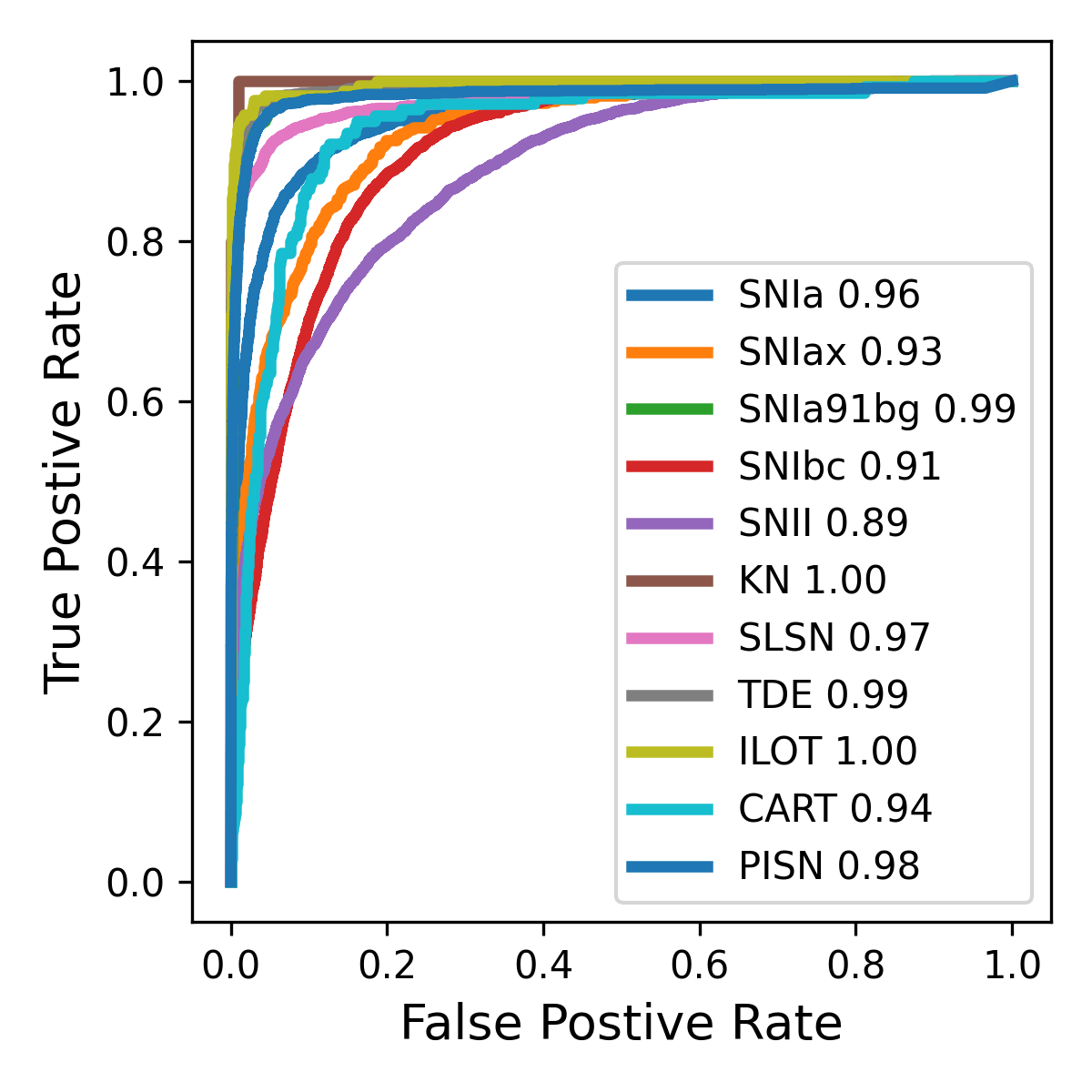}
      \caption{
ROC curves for non-recurrent transients in the \elasticc sample, illustrating the relationship between True-Positive and False-Positive rate. Area Under the Curve (AUC) scores are given in legend. Most classes, with the possible exception of SNII and SNIbc, show excellent classification scores. The KN score should be interpreted with caution, as only few simulated KNe provide the number of detections required by \parsnip. Most KN transients only caused one or two detections which passed the LSST alert generation criteria.
              }
         \label{fig:test_final_roc}
\end{figure}

\begin{figure}
   \centering
  \includegraphics[width=\hsize]{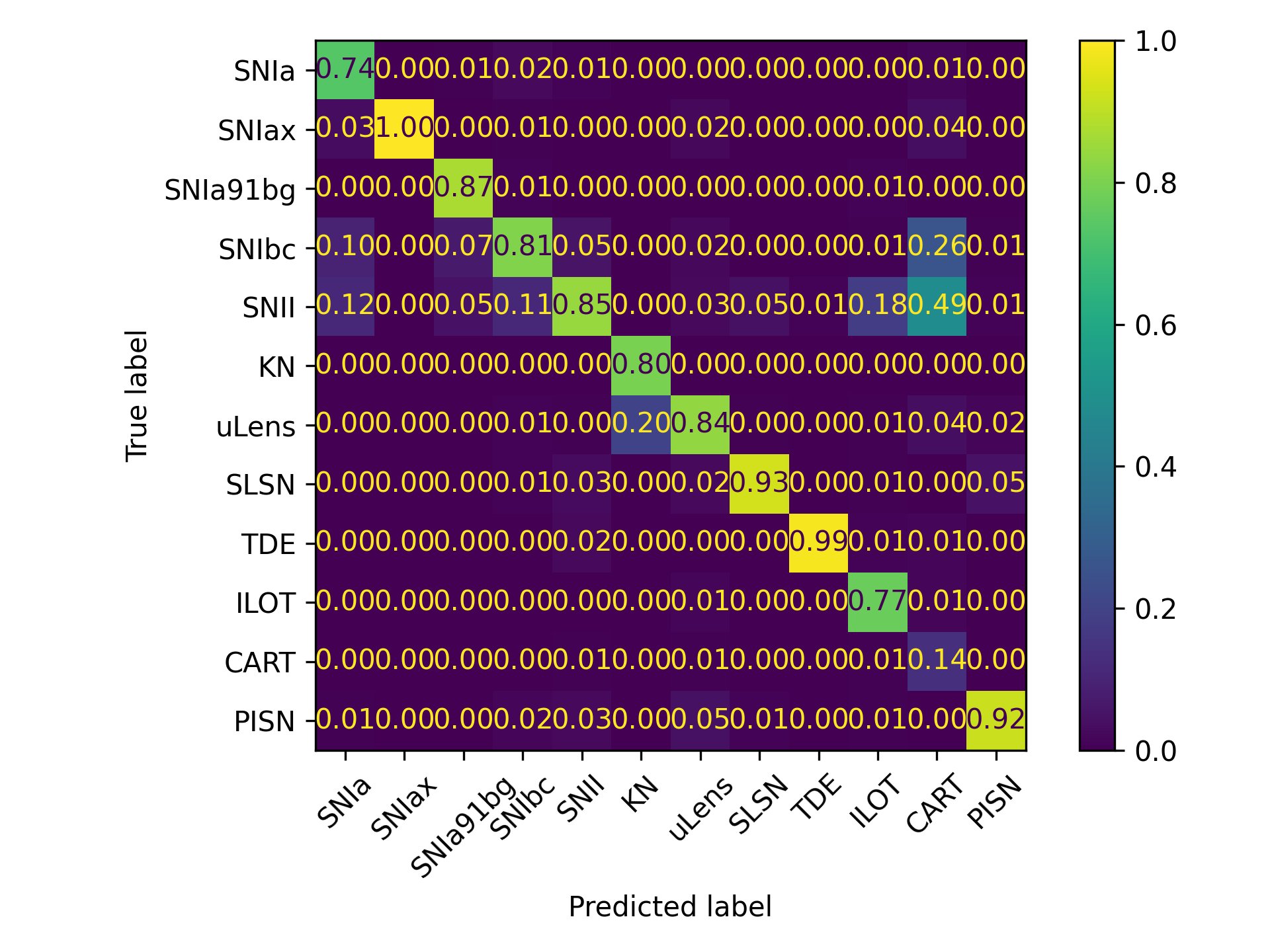}
      \caption{Confusion matrix for non-recurrent transients classified by the \final classifier. As predictions any final classification with a probability larger than $68\%$ were included, which is fulfilled for $67\%$ of all \elasticc transients. Numbers normalized along columns (predicted label).
              }
         \label{fig:test_final_cm}
\end{figure}

\begin{figure}
   \centering
  \includegraphics[width=\hsize]{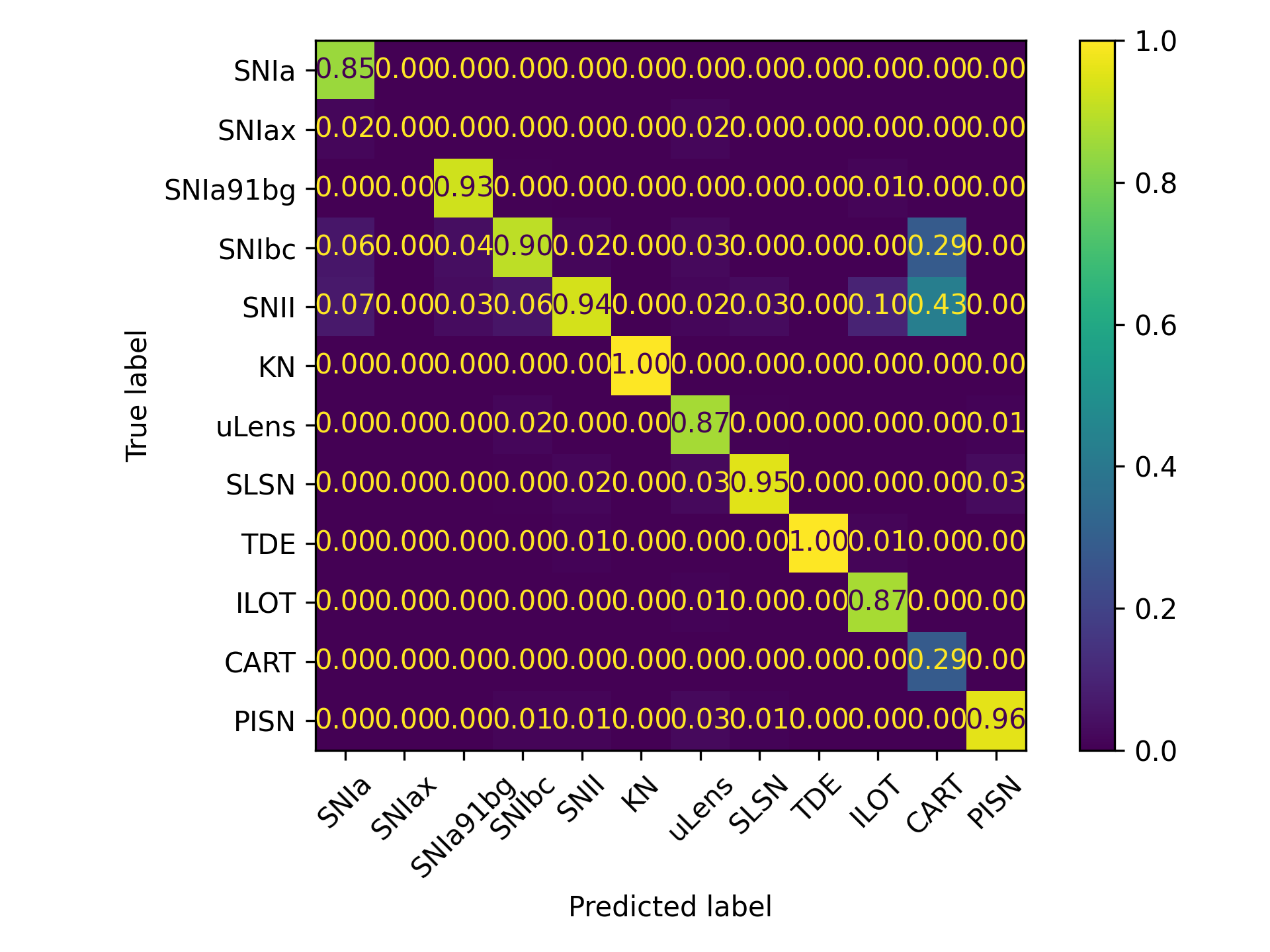}
      \caption{
      Confusion matrix for non-recurrent transients classified by the \final classifier. As predictions any final classification with a probability larger than $90\%$ were included, which is fulfilled for $43\%$ of all \elasticc transients. Numbers normalized along columns (predicted label).
              }
         \label{fig:test_final_cm2}
\end{figure}

Figures~\ref{fig:test_final_roc} to \ref{fig:test_final_cm2} show the core results of the \final classifier. These are all based on the final classification result for each transient. The first figure shows the Receiver Operating Characteristic (ROC) curve --- the relation between true-positive and false-positive rates --- for the included classes. Classifications are accurate for most classes, with Area Under the Curve (AUC) scores above $0.9$. The Kilonova (KN) class score of $1$ should, however, not be interpreted literally, since most kilonovae would produce too few detections at the LSST cadence to be processed by \parsnip.

The \newI{AUC is an integrated measurement, while an implemented astronomical program is likely to only use classifications above a certain probability threshold.} 
The confusion matrix for all classification where the classifier returns a probability of at least $0.68$ for the most likely class, which is fulfilled for $67$ \% of all transients, is shown in Figure~\ref{fig:test_final_cm}. If the classification cutoff is instead set to $0.9$, $43$ \% of alerts are classified, leading to the confusion matrix in Figure~\ref{fig:test_final_cm2}. Both show excellent classification results, typically above $80$ or even $90$ \%. The main exception seems to be a fraction of SNII and SNIbc events that are classified as CARTs.



\subsection{\elasticc blind test}

A full review of results from the streamed \elasticc blind test will be included with those of the upcoming \elasticc 2 run. However, basic evaluation metrics were calculated for \elasticc; results for the \ampel classifications are provided in Appendix~\ref{sec:blind}. Particularly interesting is the study of the phase dependence of classification, as shown in Fig.~\ref{fig:desc3}.

The same general trends as can be seen for the test sample are confirmed in the \desc run, but with somewhat worse performance. One obvious reason, added by design, is the sample priors used in \final. These were semi-arbitrarily defined, and certainly do not match the underlying class distribution that was simulated for \elasticc. Another potential cause of differences could be the alert generation process, which was regulated by \texttt{ElasticcTrainingSetLoader} for the training and internal test runs, while the LSST tools were used for the blind test evaluation. No tests were made to verify whether the output is similar.

The workflows presented here overall produced the \newI{highest precision and accuracy} classification results for extragalactic transients (the target group). 
A complete evaluation of \elasticc ($1$ and $2$) is currently in preparation by \desc.

\section{Running and extending \ampel workflows}
\label{sec:running}

The units and workflows presented above are publicly available, and can be installed and executed using the instructions contained in Listing~\ref{list:runjob}. These assume the existence of a \texttt{python} $3.10$ environment, with \texttt{poetry} and \texttt{git} available, as well as a local \texttt{MongoDB} database instance. Briefly, the steps of Listing~\ref{list:runjob} will: Clone the \texttt{Ampel-HU-astro} branch developed for \elasticc and install the required dependencies (which includes both the core \ampel libraries as well as external ML libraries such as \texttt{torch}), build a local configuration file collecting available \ampel units (\texttt{ampel\_conf.yaml}) and run a sample job which carries out classifications of $5000$ random \elasticc alerts using the workflows presented here. The sample job, \texttt{elasticc\_alerttar.yml}, is shown and discussed in more detail in appendix~\ref{sec:jobfile}.

\begin{listing}
\begin{minted}[breaklines, fontsize=\footnotesize, frame=single]{shell}
git clone --branch elasticc2 https://github.com/AmpelAstro/Ampel-HU-astro.git
cd Ampel-HU-astro/
poetry install -E "ztf sncosmo extcats notebook elasticc"
ampel config build -out ampel_conf.yaml >& ampel_conf.log
ampel job --config ampel_conf.yaml --schema examples/elasticc_alerttar.yml
\end{minted}
\caption{Installing local \ampel environment and running the \elasticc sample job.}
\label{list:runjob}
\end{listing}

Many opportunities exists after running the sample job: Filter or selection functions can be edited to finetune the alert selection, improved \texttt{Parsnip} or \texttt{XGBoost} models can be directly applied or new kinds of ML models be introduced by forking \texttt{T2RunParsnip} or \texttt{T2MultiXGBClassifier} and adapt it to another method for applying models --- the \ampel IO class structure ensures that the model will be provided the required data. Workflows updated in this way can be applied to larger \elasticc alert samples, included as part of a software distribution or join upcoming \desc data challanges. In a few years, the full LSST alert stream will be applied to a series of such workflows hosted by the \ampel live instance at DESY.

\section{Conclusions}
\label{sec:conc}

The future of time-domain astronomical research is exciting, but making full use of current and upcoming facilities requires powerful and flexible tools. We have here presented three blueprints for machine-learning-based workflows designed for the \elasticc simulation of LSST alerts. While sharing some components, each was designed to highlight different science goals:

\begin{itemize}
    \item \snguess can correctly select young extragalactic transients for followup ($99$\% true-positive rate). AGN outbursts are a contaminant, but can be removed either by avoiding galaxy cores or adding further selection logic. \snguess can easily be combined with external multi-messenger constraints to produce pure streams of counterpart candidates.
    \item The \follow workflow shows how an alert stream can be sampled in order to create labeled subsets designed for verifying or improving photometric classification algorithms. We discuss a schema for how to understand the potential confusion among SNe Ia and SNe Ibc for cosmological distance estimates.
    \item \final adds explicit priors to the classification output, based on information external to the pure transient lightcurves. Several such information sources exist, and should be taken into account for optimal performance. We here include priors based on candidate redshift, host environment and the expected relative rates. Classes achieve AUC scores above $0.9$ and for  "trustworthy" classification ($prob>0.68$) the predicted class is correct at a rate well above $80$ $\%$ for most transient types.
\end{itemize}

All of these have been implemented as jobs running in the open source \ampel processing framework. This allows both for scalability --- jobs can be run in parallel at computer centers --- as well as reproducibility and modularity. The full classification pipelines can be executed in a local development mode. In particular, this allows individual science groups to optimize workflows for their particular science goals. The list of potential changes is long: from how alerts are filtered, to which classification algorithms are employed (off the shelf algorithms were used here), how these results are parsed and combined, what priors are used and finally which thresholds are set for reactions and how these should look. The sample workflows include several components designed for use during the LSST survey rather than the \elasticc simulation (for example the redshift sampling), such that they can directly be used as inspiration for the design of VRO era science exploration.

Interested users are invited to set up a local \ampel environment and run the included \elasticc classification sample job. A live \ampel instance will be one of the endpoints for the LSST alert streams, and general labels will be available based on these workflows within minutes of alert generation. Developers with ideas of how to create even better classification schema can use the provided workflows as starting points for this work. Once completed, the standardized format will allow the improved pipelines to be shared, put to live usage on real-time streams processed by the \ampel instances (for instance ZTF, LSST or LS4) or to join upcoming \elasticc classification challenges. Questions regarding any of these steps can be directed to the corresponding authors.

Finally, \ampel makes it straightforward to create real-time workflows which uphold F.A.I.R. principles: jobfiles are self-contained and can be referenced and made available through public channels, they can be directly executed by other scientists and used as starting points for extended, updated or improved versioned. Simultaneously, all actions and results derived while executing a workflow are logged and recorded by the provided \ampel reproducibility features.

\begin{acknowledgements}
      \newR{
      We are very grateful for the work in setting up and executing the ELAsTiCC challenge carried out by the LSST/DESC Team.
      }
\end{acknowledgements}

%
%

\bibliographystyle{aa} 
\bibliography{ampelMLref.bib}

\begin{appendix} 

\section{Unit summary}
\label{sec:units}

The purpose and logic of each of the units employed by the \snguess, \follow and \final channels (Figures \ref{fig:snguessflow} and \ref{fig:parsnipflow}) is provided here. The full implementation available in the Ampel-ZTF\footnote{https://github.com/AmpelProject/Ampel-ZTF} or Ampel-HU-astro\footnote{https://github.com/AmpelProject/Ampel-HU-astro} repositories (in a \texttt{python} file under the unit name).

\paragraph{ElasticcDirAlertLoader} Parse a local directory containing LSST alerts as simulated by \elasticc 1, iterating through each alert.

\paragraph{ElasticcTrainingSetLoader} The \elasticc training sample was distributed as a set of full lightcurves saved in \texttt{SNANA} format. Each set (pair of header and body) contains lightcurves simulated using a single transient model. This unit will read one such set and split into single alerts as expected from a real-time survey. For each lightcurve, each new photometry point tagged as being above the detection threshold will yield a new alert (containing older datapoints as previous detections).

\paragraph{ReallySimpleLSSTFilter} The core part of the first \ampel tier (T0) is to select or filter which alerts might be interesting and queue these for further study. The \texttt{ReallySimpleLSSTFilter} is, as suggested, a simple filter which selects transients based on the number of detection and the total alert duration. Within \elasticc we know all alerts to be interesting, and this filter is thus configured to pass \emph{all} alerts through. A more realistic filter will base selections on existing alert properties or through matching with external catalogs, such that only transients of potential interest are saved.

\paragraph{T2GetDiaObject} A simple unit which propagates the \elasticc alert identification to the subsequent publishing unit.

\paragraph{T2ElasticcRedshiftSampler} Key information when selecting or analysing transients comes from determining the potential host galaxy (if believed to be extragalactic) as well as the distance/redshift. The \texttt{T2ElasticcRedshiftSampler} unit extracts these properties from data provided in the \elasticc alerts. For a fully developed LSST stream this job would consist both in collecting information from external catalogs and parse information provided in the alerts, and then perform any operation related to extracting photometric redshifts or sampling from these.  The simulated \elasticc alerts introduces much of these complications as information is provided for zero, one or two potential host galaxies, together with photometric redshift information given as quantiles.
\texttt{T2ElasticcRedshiftSampler} uses this information as follows: The most likely host galaxy is chocen as the one with the smallest $\texttt{snsep}/\texttt{sqradius}$ ratio. This is based on the assumption that also a more distant galaxy (larger \texttt{snsep}) can be the likely host if sufficiently large (\texttt{sqradius}).
The unit will then sample the photometric redshift uncertainty belonging to the host galaxy and return a set of discrete redshifts together with probability weights. The default behaviour is to provide three samples with suggested weights $0.2$, $0.6$ and $0.2$. These redshifts are chosen as the mean of the photometric redshifts at the \{${0, 0.1, 0.2, 0.3, 0.4}$\}, \{${0.3, 0.4, 0.5, 0.6, 0.7}$\} or \{${0.6,0.7,0.8,0.9,1.0}$\} quantiles, respectively for the three samples. These weights and quantiles, while arbitrarily chosen, shows how one can try to combine a possibly irregular photometric redshift distribution into a set of values which spans a reasonable range. Subsequent photometric classification units will be run at each of the redshift samples, and the goodness of fits are combined with the probability weights to determine the best fit global redshift.

Alerts without any reports of nearby host galaxies are likely to be stellar variables, but a fraction could also be seemingly hostless extragalactic transients. To simulate this, events without any identified host galaxies get redshift samples placed at \{$0.01, 0.2, 0.5, 0.8$\} with weights \{${0.4, 0.2, 0.2, 0.2}$\}. Finally, this unit will also propagate the ($u-g$) and ($u-i$) host galaxy colors.

Note that this unit was developed more to highlight the kind of operations that a live LSST program might wish to carry out than being optimized for \elasticc. The redshift sampling methodology chosen is non-optimal for the symmetric photometric uncertainties of \elasticc 1, and the conservative treatment of galactic transients means that a large fraction of these will be inefficiently processed as potential extragalactic events.

\paragraph{T2RiseDeclineStat} Calculates up to $32$ features based on photometry and alert information tied to the transient state. The features were selected to align with measurables traditionally examined shortly after a young transients has been discovered and most are designed such that only a minimum of datapoints are needed. These features range from temporal ranges (duration, time gap to last significant upper limit), boolean estimates (can a peak be seen, was a rise observed?, is it rising?), magnitudes (at detection, at peak, etc), colors (at detection, peak and last phase) and linear slope coefficients (rise, decline). A full description of features and the methodology can be found in \cite{2022A&A...665A..99M}.

\paragraph{T2MultiXGBClassifier} Features, as calculated by \texttt{T2RiseDeclineStat}, are ideal for use as input to binary classification trees, a method first presented and verified based on real ZTF alerts \citep{2022A&A...665A..99M}. The \texttt{T2MultiXGBClassifier} is designed to apply an input series of classifiers trained using the gradient boosting technique as implemented in the \texttt{XGBoost} library, where the features determined by the parent class (\texttt{T2RiseDeclineStat}) are used as input. The unit will, for each loaded tree, return the estimated probability that the hypothesis is correct. For the models trained here, this corresponds to the candidate being a non-recurrent transient.

The structure of \elasticc led to the following two modifications compared with the \cite{2022A&A...665A..99M} version: First, the redshift, redshift error and host separation are retrieved from the output of the \texttt{T2ElasticcRedshiftSampler} and added to the list of features. Second, the base feature calculation only accepts \emph{positive} detections as real as it was initially developed for supernova (positive) detections. Transient classes such as AGNs, \ulens and eclipsing binaries will frequently display mainly negative detections if the baseline (reference) was determined in an excess state. To reflect this known inconsistency we add the following classification conditions for transients with no significant positive detections:
\begin{itemize}
    \item An alert with multiple significant negative detections and $z>0.001$ will be labeled as AGN.
    \item Alerts at $z<0.001$ and with more than $20\%$ of all photometric datapoints being negative are labeled as \ulens events.
    \item Alerts at $z<0.001$ and with less than $20\%$ of all photometric datapoints being negative are labeled as eclipsing binary events.
    \item Any other alert will be labeled as "Noise". Although the ELASTiCC simulation does not contain pure noise, this is likely to dominate this category for real data streams.
\end{itemize}

\paragraph{T2RunParsnip} Classifies a given input lightcurve, positioned at a set of discrete redshifts, using a specified \texttt{Parsnip} model. The classification predictions at each redshift are combined to a single set of probabilities based on the $\chi^2$ model fit values at each redshift and incorporating the weights provided by \texttt{T2ElasticcRedshiftSampler}.
The final classification will thus depend on the model fit at each of the sampled redshift, such that a good fit can compensate for a possibly catastrophic redshift estimate. Several ML models, including \parsnip, have modes where distance is included as one of the fit parameters. However, predictions are typically not rapidly varying with distance while including distance often increases execution times. Running at discrete redshifts provides a middle ground, where the number of samples can be set to vary for example with redshift origin.
A relatively common example will be extragalactic transients with no detected host galaxy, for which a ML model at a common LSST redshift (e.g. $\sim 0.4$) can allow for a much better fit compared with the same model at $z \sim 0$.

\paragraph{T2ElasticcReport} This unit will combine data from the classifiers into reports following the expected \elasticc schema for broker replies, creating one report for each classifier. For the \final classifier this includes applying the priors presented above.

\paragraph{ElasticcClassPublisher} Posts the classification report to the dedicated server maintained by \desc.

\paragraph{ElasticcClassTablePrinter} Collects and exports \elasticc classifications from the DB to an external file. Unit parameters regulates which classifiers are to be included, and whether a minimal probability should be required for inclusion.

\section{Results: ELAsTiCC blind test}
\label{sec:blind}

As was mentioned above, the ELAsTiCC project was setup to receive the classifications of the still blinded alerts from the broker teams.
We here present and discuss the evaluation of the above AMPEL models which were done as part of the continous ELAsTiCC evaluation. Figures were downloaded from the ELAsTiCC project page\footnote{\url{https://desc-tom.lbl.gov/elasticc/}}. Note that while the training sample was generated to create roughly balanced samples between the different classes, the (masked) test sample was generated based on still unknown redshift and rate distributions. Differences are thus to be expected and directly illustrates the challenge of optimally using the deep VRO photometry.

\begin{figure*}
\center{ \adjustbox{trim=10cm 0 0 0}{
\includesvg[inkscapelatex=false,height=0.9 \textheight] {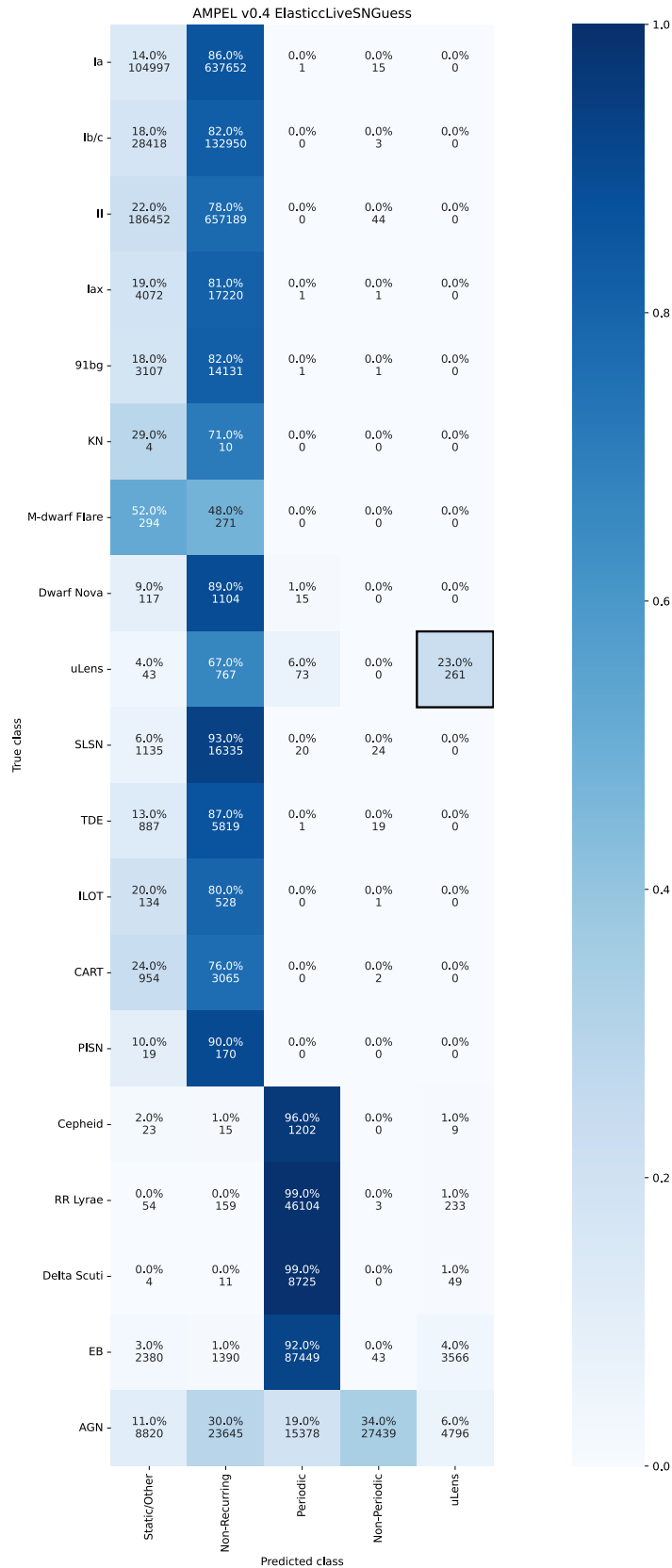} }
}
\caption{
   Confusion matrix generated by \desc based on the blinded alert stream of \elasticc for the \snguess classifier. The main goal was to disginguish non-recurrent alerts from recurring (periodic or non-periodic). The Static/Other label was returned when \snguess did not find sufficient features to complete a classification. Classifications, when provided, were correct for almost all alerts. The AGN and \ulens classes perform worse, as discussed in the text.
}
\label{fig:desc2}
\end{figure*}

The confusion matrix for the \snguess channel is shown in Fig. ~\ref{fig:desc2}. The \emph{Static/Other} classification was evoked for cases where \snguess did not produce a final evaluation, most often occuring due to no significant detection in any of the VRO g, r, i or Z bands. Alerts which do pass this significance threshold are for most classes correctly classified as being either non-recurring or periodic. There are two main exceptions to this trend in $\mu$lens and AGN events, which show worse behaviour. This is in both cases due to a large fraction of significant but negative detections, which the original \texttt{RiseDecline} features were not designed to handle well but which has been corrected for in recent unit updates. Whether the threshold for the required signal strength should be lowered is less certain as more classifications would be returned but with a lower quality. The optimal choice here will depend on the full scientific program and will have to be tuned based on immediate goals and follow-up resources. The \snguess channel was designed specifically to select young transients for follow-up with high purity, goals which are well fulfilled. Similarly, contamination from AGN could be drastically reduced through only accepting transients offsets from the core of galaxies. Whether this increase in purity is worth the decreased completeness will vary between applications.

\begin{figure*}
\includesvg[inkscapelatex=false,width=1. \hsize]{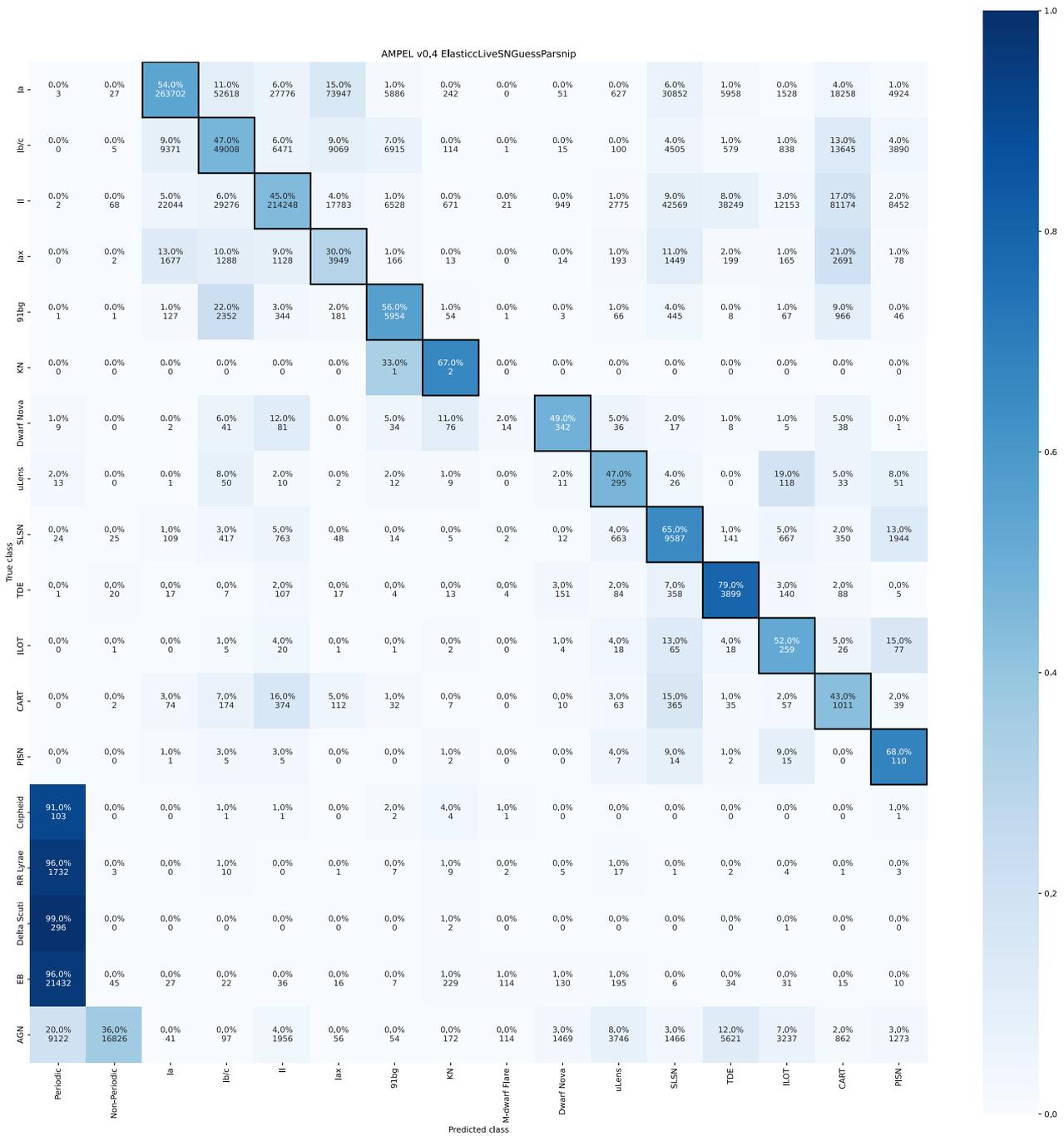}
\caption{
Confusion matrix for blind evaluation of \elasticc test sample for the \follow classifier ("SNGuessParsnip"). Recurring events, not the classifier target, are consistently classified as such. True classification fractions, normalized to the true class generally vary between $45$ and $70$\%. TDEs are most accurately classified ($79$ \%) and SNIax being the least ($30$\%).
}
\label{fig:descfollow}
\end{figure*}

\begin{figure*}
\includesvg[inkscapelatex=false,width=1. \hsize]{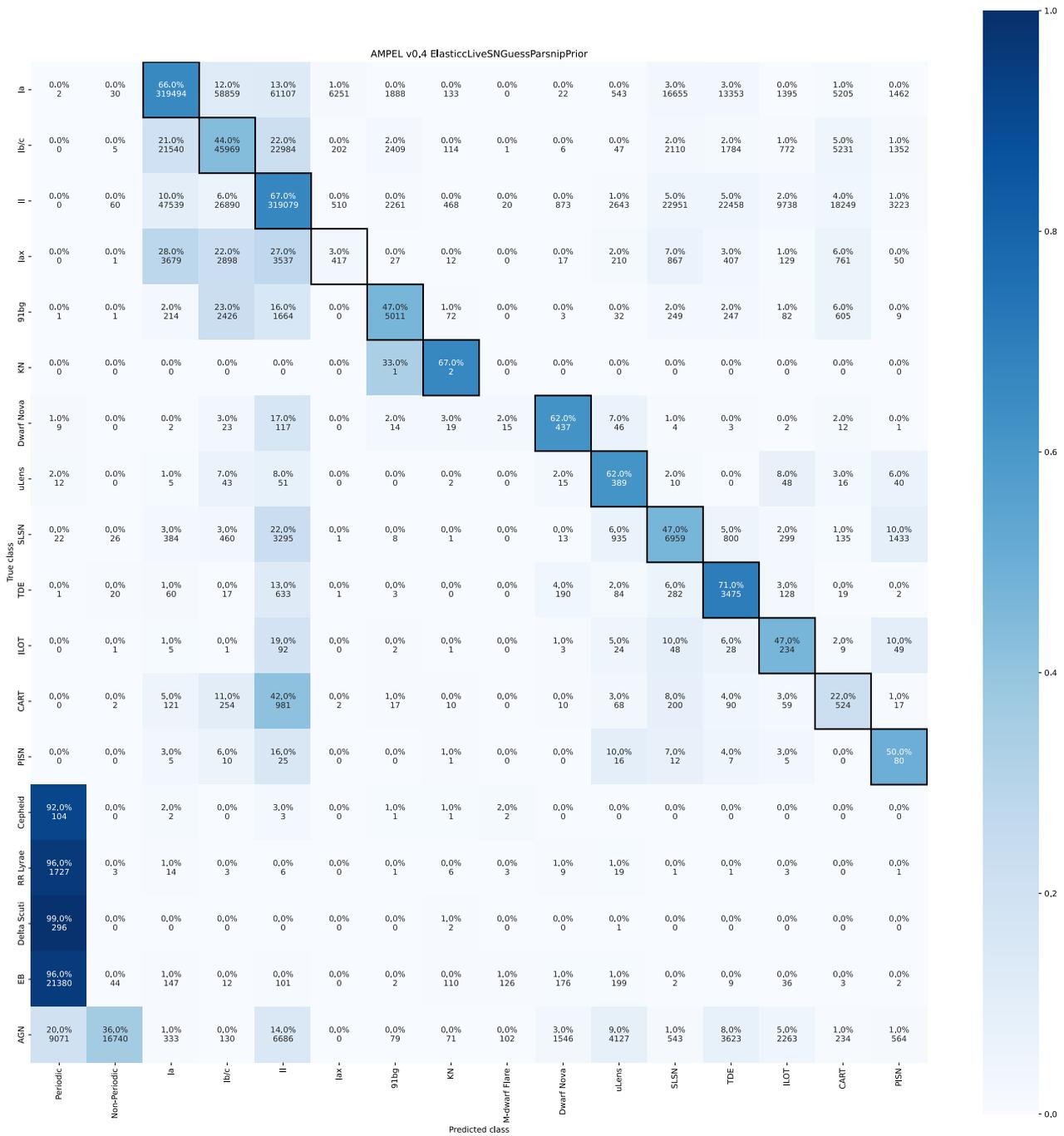}
\caption{Confusion matrix for blind evaluation of \elasticc test sample for the \final classifier ("SNGuessParsnipPrior"). Recurring events, not the classifier target, are consistently classified as such. True classification fractions, normalized to the true class generally vary between $50$ and $70$\%. The applied priors shift probabilities as expected with respect to \follow (Fig above): Common SN types, like SNIa, are more consistently correctly classified at the expense of rare transient types like SNIax.
}
\label{fig:descfinal}
\end{figure*}

The full confusion matrices for the \follow and \final channels are shown in Figs.~\ref{fig:descfollow} and~\ref{fig:descfinal}. Overall these results match those derived from the testset withhold from the training data, but with a decreased performance. Variable stars and AGN derive their behaviour from the \snguess channel (as \texttt{PARSNIP} was not trained on these), while no attempts at classifing M-dwarf stellar outbursts was made. The overall classification performance for the \follow channel varies between $30\%$ and $80\%$ depending on the full transient class. The intraclass confusion is behaving as expected: SNe I alerts (for example) are most frequently misclassified as SNIb/c or SNIax events. Looking at \final results, adding the fairly arbitrary priors still improved results significantly with most transient types having true classification probabilities above $60\%$. Several rare cases show significantly worse outcome, including SNIax and CART events. Clearly the priors added were incorrect, causing a bias in the final classifications. We also note that two out of three KN events distributed during this segment of ELAsTiCC were correctly identified.

\begin{figure}
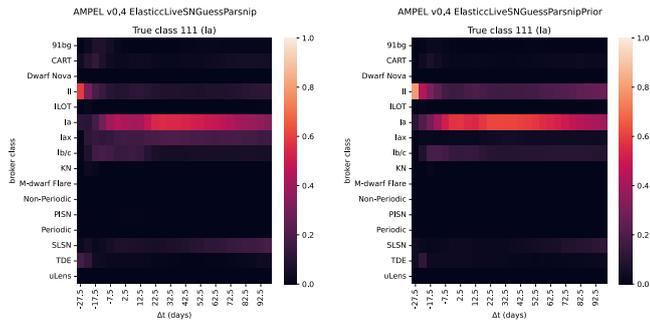

\includesvg[inkscapelatex=false,width=0.48 \hsize]{figures/88_111_ampel04_parsnip.svg}
\includesvg[inkscapelatex=false,width=0.48 \hsize]{figures/89_111_ampel04_parsnipprior.svg}
\caption{Classification precision for SNe Ia vs transient phase (time relative to peak). SNe Ia are generally correctly classified after a phase of $-7.5$. Very early detections are often mistakenly classified as SNII. There is a low level confusion with the very similar SNIax class, as
 well as with SNIbc around peak. The \final classifier mode (right) has a more accentuated trend compared with \follow (left).
}
\label{fig:desc3}
\end{figure}

The above results combined alerts from all transient phases as well as across all redshifts, while we expect classification performance to vary strongly as these vary. As an example, figure~\ref{fig:desc3} illustrates the time dependence of SNIa classification for the \follow and \final channels. Several interesting things are seen here. Alerts with extremely early or late phases compared with the main SNIa population are routinely misclassified as SNII or SLSN (respectively). At the same time alerts at what corresponds to post-peak phases of SNIa, including the second bump, show much stronger classification results than the aggregated values seen in Fig.~\ref{fig:descfinal}.

\section{The \ampel \elasticc jobfile}
\label{sec:jobfile}

The workflow encoded in \texttt{elasticc\_alerttar.yml} can be summarized through these steps:
\begin{itemize}
    \item Download a set of $5000$ \elasticc alerts and ingest these into a local \ampel collection.
    \item Execute supporting units to sample the photometric redshift uncertainty.
    \item Download and execute the ML models used in \texttt{T2MultiXgbClassifier} and \texttt{T2RunParsnip} for each alert.
    \item Collect the derived information and based on these produce \snguess, \follow and \final classifications.
    \item Output all classification probabilities (following the \elasticc taxonomy) larger than $5$\% into a local file (\verb|elasticcClassifications.csv|).
\end{itemize}

\section{Sample \ampel DB entries}
\label{sec:dbentries}

Listings\xspace\ref{list:stock} to \ref{list:t2} show sample \ampel DB documents which will be created during the processing of the \elasticc sample job.

\begin{listing}[h]
\begin{minted}[fontsize=\footnotesize]{yaml}
    {"_id" : ObjectId("650335ab82ec01e3aa15ffc4"),
    "stock" : 44361271,
    "channel" : [
        "Elasticc"
    ],
    "journal" : [
        {
            "ts" : 1694709163,
            "run" : 1,
            "tier" : 0,
            "channel" : "Elasticc",
            "alert" : 88722542042,
            "action" : 335610113,
            "traceid" : {
                "alertconsumer" : 3699778560255607892,
                "shaper" : 4942518040143172695,
                "muxer" : 1497822671722126522,
                "combiner" : 4699038342887496180
            },
            "upsert" : [
                88722542042,
                ...
            ],
            "link" : 1395918359
        },
        {
            "tier" : 2,
            "ts" : 1694709203,
            "process" : "elasticc-nov#1",
            "run" : 2,
            "action" : 262144,
            "channel" : [
                "Elasticc"
            ],
            "unit" : "T2ElasticcRedshiftSampler",
            "doc" : { "$binary" : "ZQM1q4LsAeOqFf1P", "$type" : "00" },
            "traceid" : {
                "t2worker" : -4535828744713078706,
                "t2unit" : -3918940777486538662
            }
        },
        ...
    ],
    "tag" : [
        "LSST"
    ],
    "ts" : {
        "Elasticc" : {
            "tied" : 1694709163,
            "upd" : 1694710655
        },
    }
    }
\end{minted}
\caption{Sample \ampel \texttt{stock} document. Each (presumed) unique object is assigned a stock id, and the stock collection records which \texttt{channels} are associated to the object as well as what operations were carried out (the \texttt{journal})}
\label{list:stock}
\end{listing}

\begin{listing}[h]
\begin{minted}[fontsize=\footnotesize]{yaml}
    {"_id" : ObjectId("650335ab82ec01e3aa15fdbc"),
    "id" : 88722542042,
    "body" : {
        "diaSourceId" : 88722542042,
        "ccdVisitId" : -1,
        "diaObjectId" : 44361271,
        "parentDiaSourceId" : null,
        "midPointTai" : 60563.1116,
        "filterName" : "i",
        "ra" : 317.191946981264,
        "decl" : -0.704240902148387,
        "psFlux" : -67431.9765625,
        "psFluxErr" : 1011.32653808594,
        "snr" : -66.6767578125,
        "nobs" : 279.0
    },
    "channel" : [
        "Elasticc"
    ],
    "meta" : [
        {
            "ts" : 1694709163,
            "run" : 1,
            "alert" : 88722542042,
            "activity" : [
                {
                    "action" : 2048,
                    "channel" : "Elasticc"
                },
                {
                    "action" : 64,
                    "tag" : "LSST"
                }
            ],
            "traceid" : {
                "shaper" : 4942518040143172695
            }
        }
    ],
    "stock" : [
        44361271
    ],
    "tag" : [
        "LSST",
        "LSST_I",
        "LSST_DP"
    ] }
    \end{minted}
\caption{Sample \ampel \texttt{t0} document. Each measurement is recorded as a \texttt{photopoint} in the t0 collection. The photopoint body is immutable, and only activity connected to the object is recorded in the \texttt{meta} field.
}
\label{list:t0}
\end{listing}

\begin{listing}[h]
\begin{minted}[fontsize=\footnotesize]{yaml}
    {"_id" : ObjectId("650335ab82ec01e3aa15ffd7"),
    "link" : 1395918359,
    "stock" : 44361271,
    "channel" : [
        "Elasticc"
    ],
    "dps" : [
        44361271,
        88722542042,
        ...
    ],
    "meta" : [
        {
            "run" : 1,
            "ts" : 1694709163,
            "tier" : 0,
            "alert" : 88722542042,
            "code" : 0,
            "activity" : [
                {
                    "action" : 133136,
                    "channel" : "Elasticc"
                },
                {
                    "action" : 64,
                    "tag" : "LSST"
                }
            ],
            "traceid" : {
                "alertconsumer" : 3699778560255607892,
                "combiner" : 4699038342887496180,
                "muxer" : 1497822671722126522
            }
        }
    ],
    "tag" : [
        "LSST"
    ]
}
\end{minted}
\caption{Sample \ampel \texttt{t1} document. The \texttt{state}, i.e. information or datapoints known by a user (channel) at some point in time is recorded in the t1 collection.
}
\label{list:t1}
\end{listing}

\begin{listing}[h]
\begin{minted}[fontsize=\footnotesize]{yaml}
{
    "_id" : ObjectId("650335ab82ec01e3aa15fd53"),
    "config" : 9100365094983736939,
    "link" : 1395918359,
    "stock" : 44361271,
    "unit" : "T2RunParsnip",
    "channel" : [
        "Elasticc"
    ],
    "code" : 0,
    "meta" : [
        ...
        {
            "run" : 7,
            "ts" : 1694709311,
            "tier" : 2,
            "code" : 0,
            "duration" : 3.448,
            "activity" : [
                {
                    "action" : 139264
                }
            ],
            "traceid" : {
                "t2worker" : 1693292732686570277,
                "t2unit" : 4967151258208453945
            }
        }
    ],
    "tag" : [
        "LSST"
    ],
    "body" : [
        {
            "model" : "model1-30pct-sn+2ulens+dwarfs-mw.h5",
            "classifier" : "model1-30pct-sn+2ulens+dwarfs-mw-aug10.pkl",
            "abort_maps" : {},
            "jdstart" : -Infinity,
            "jdend" : Infinity,
            "z" : [ 0.01, ... ],
            "z_source" : "default",
            "z_weights" : [ 0.4, ],
            "predictions" : {
                "0.01" : {
                    "ra" : NaN,
                    "dec" : NaN,
                    "type" : "Unknown",
                    "redshift" : 0.01,
                    "parsnip_reference_time" : 2460495.79823479,
                    "parsnip_scale" : 174460.953125,
                    "reference_time" : 2460503.4783524,
                    "reference_time_error" : 0.0483041293919086,
                    "color" : -0.11637556552887,
                    "color_error" : 0.0118869617581367,
                    "amplitude" : 298.269073486328,
                    "amplitude_error" : 5.91950988769531,
                    "s1" : 1.24571561813354,
                    "s1_error" : 0.0538764894008636,
                    "s2" : -0.749227523803711,
                    "s2_error" : 0.0589446611702442,
                    "s3" : -0.455027937889099,
                    "s3_error" : 0.0140000898391008,
                    "total_s2n" : 476.459411621094,
                    "count" : 43,
                    "count_s2n_3" : 21,
                    "count_s2n_5" : 21,
                    "count_s2n_3_pre" : 0,
                    "count_s2n_3_rise" : 5,
                    "count_s2n_3_fall" : 16.0,
                    "count_s2n_3_post" : 0,
                    "model_chisq" : 227158.59375,
                    "model_dof" : 37,
                    "luminosity" : -10.5354884851385,
                    "luminosity_error" : 0.0215505678206682,
                    "chi2pdf" : 0.0
                },
                ...
            },
            "classifications" : {
                "0.01" : {
                    "CART" : 0.0273714825568351,
                    "ILOT" : 0.0219947098607236,
                    "KN" : 0.00266194307622392,
                    "Mdwarf-flare" : 0.0239980869801839,
                    "PISN" : 0.00125751193039502,
                    "SLSN-I" : 0.00190615907139106,
                    "SNII" : 0.017974263385055,
                    "SNIa" : 0.032107158172344,
                    "SNIa91bg" : 0.0136321523174439,
                    "SNIax" : 0.0471953413102884,
                    "SNibc" : 0.0235590438320758,
                    "TDE" : 0.0292363558426192,
                    "dwarf-nova" : 0.416904035806487,
                    "uLens" : 0.340201755857934
                },
                ...
            },
            "Failed" : "NoFit"
        }
    ]
}
\end{minted}
\caption{Sample \ampel \texttt{t2} document. The output of each t2 analysis unit, in this case \texttt{T2RunParsnip} is recorded in the \texttt{body} field in the t2document associated to the object state and unit configuration. }
\label{list:t2}
\end{listing}

\end{appendix}

\end{document}